\documentclass[preprint,preprintnumbers,aps,pra,amsmath]{revtex4-1}

\usepackage[T1]{fontenc}

\usepackage{color,graphicx,cases,enumitem}

\usepackage[customcolors]{hf-tikz}
\tikzset{style grey/.style={
    set fill color=black!30!white!70,
    set border color=white,
  },
  hor/.style={
    above left offset={-0.15,0.43},
    below right offset={0.2,-0.19},
    #1
  }
}

\newcommand*{\ket}[1]{\left|#1\right>}

\begin{document}

\title[Formulating Complex Band Structure]{A Unified Perspective of Complex Band Structure: Interpretations, Formulations, and Applications}
\author{Matthew G.\ Reuter}
\email{matthew.reuter@stonybrook.edu}
\affiliation{Department of Applied Mathematics \& Statistics and Institute for Advanced Computational Science, Stony Brook University, Stony Brook, New York 11794, United States}
\date{\today}

\begin{abstract}
Complex band structure generalizes conventional band structure by also considering wavevectors with complex components. In this way, complex band structure describes both the bulk-propagating states from conventional band structure and the evanescent states that grow or decay from one unit cell to the next. Even though these latter states are excluded by translational symmetry, they become important when translational symmetry is broken via, for example, a surface or impurity.

Many studies over the last 80 years have directly or indirectly developed complex band structure for an impressive range of applications, but very few discuss its fundamentals or compare its various results. In this work we build upon these previous efforts to expose the physical foundation of complex band structure, which mathematically implies the existence of complex band structure. We find that a material's static and dynamic electronic structure are both completely described by complex band structure. Furthermore, we show that complex band structure reflects the minimal, intrinsic information contained in the material's Hamiltonian. These realizations then provide a context for discussing, comparing, and unifying the different formulations and applications of complex band structure that have been developed over the years. Ultimately, this discussion introduces the idea of examining the amount of information contained in a material's Hamiltonian so that we can find and exploit the minimal information necessary for understanding a material's properties.
\end{abstract}

\maketitle

\section{Introduction}
\label{sec:introduction}
Band structure has become a critical component of condensed matter physics because it provides information on a crystalline material's electronic, magnetic, and optical properties \cite{bk:ashcroft-1976}. Materials that are mostly, but not perfectly, crystalline are also of interest, but band structure no longer provides a complete characterization of these materials due to disorder (\textit{e.g.},\ an impurity or surface). Complex band structure (CBS) \cite{goodwin-205-1939, kohn-809-1959, heine-300-1963, heine-1-1964, krieger-776-1967, prodan-035128-2006} generalizes this conventional band structure to account for disorder by considering wavevectors with complex components. Although CBS has been derived and developed in numerous contexts (\textit{vide infra}), the key physical and mathematical structures that underly CBS have gone unrecognized. Consequently, CBS's generality has not been fully realized and some CBS results are known only to the specific communities for which they were derived. In this work we expose the key assumptions behind CBS and then use them to show that CBS completely describes a material's static and dynamic electronic properties. We also interpret CBS as the minimal amount of information to accomplish this. In a broader sense, our analysis examines the information content of a material's Hamiltonian so that this information can be exploited to characterize the material's properties. With this interpretation of CBS, we proceed to discuss CBS's applications and unify its many results.

We accomplish this as follows. The remainder of this section provides additional background on CBS, including some of its history and applications. We proceed to introduce a concrete example in section \ref{sec:example}, which illustrates the fundamental properties of CBS. Our primary contribution comes in section \ref{sec:unified}, where we develop our interpretation of CBS by considering the information content of a material's Hamiltonian. This analysis exposes the underlying physical and mathematical structures that imply the existence of CBS. It is also performed at the level of operators, rather than that of matrices and basis sets, thereby eliminating possible ambiguities about basis-set effects. Next, sections \ref{sec:cbs-wf}--\ref{sec:extended-coupling} review existing results of CBS using a consistent nomenclature and notation. Finally, section \ref{sec:conclusions} summarizes our results and speculates on future directions.

\subsection{Background}
\label{sec:introduction:background}
At its core, conventional band structure is rooted in the translational symmetry of a crystalline material, where periodicity of the potential allows invocation of Bloch's theorem. We thus consider wavefunctions of the form
\begin{equation}
\psi_{n, \vec{k}} (\vec{r}) = e^{i \vec{k}\cdot \vec{r}} u_{n, \vec{k}}( \vec{r} ),
\label{eq:bloch-theorem}
\end{equation}
where $\vec{k}$ is a wavevector (usually in the first Brillouin zone), $n$ is the band index, and $u$ is a function with periodicity of the lattice. When used in the Schr\"odinger equation, $\vec{k}$ can be treated as a parameter (\textit{i.e.},\ each $\vec{k}$ produces an independent system), and the eigenstates for a given $\vec{k}$ are calculated using standard techniques. The ensuing dispersion relation for the $n$\textsuperscript{th} band, $E_n(\vec{k})$, details the material's band structure.

Despite the numerous applications of this conventional band structure, many systems of interest lack perfect translational symmetry. Surfaces and interfaces, for instance, break translational symmetry in (at least) one direction. These systems still possess a ``repeat unit'' similar to the periodic unit cell, and Eq.\ \eqref{eq:bloch-theorem} may remain applicable if solutions with complex $\vec{k}$ are also admitted \cite{kohn-809-1959, heine-300-1963, heine-1-1964, krieger-776-1967}.

Imaginary components of $\vec{k}$, when present, indicate evanescence; that is, the state grows or decays in magnitude from one repeat unit to the next. These states have sometimes been called ``generalized Bloch functions''. Conversely, bulk states that propagate throughout the material have real $\vec{k}$. For physicality, we restrict our attention to complex wavevectors that describe states with real energies \cite{heine-300-1963, heine-1-1964, krieger-776-1967}, and the set of such $\vec{k}$ constitutes the material's complex band structure (CBS). In other words, CBS extends the dispersion relation to complex $\vec{k}$, such that it describes both the propagating bulk states from conventional band structure and also any evanescent states that are forbidden by translational symmetry \cite{heine-1-1964}.

Given CBS's broad applicability, it has found use in myriad contexts: %
Surfaces and interfaces \cite{maue-717-1935, goodwin-205-1939, shockley-317-1939, heine-1-1964, feuchtwang-731-1967, marcus-925-1968, garcia-moliner-1789-1969, garcia-moliner-1797-1969, pendry-59-1970, kalkstein-85-1971, jepsen-416-1971, appelbaum-2166-1972, cottey-2583-1972, appelbaum-4973-1974, rehr-448-1974, pendry-87-1975, appelbaum-479-1976, bross-173-1977, allen-917-1979, allen-1454-1979, dy-4237-1979, brasher-4868-1980, lee-4988-1981, lee-4997-1981, lee-355-1981, dy-633-1982, chang-605-1982, chang-3975-1982, kambe-443-1982, schulman-2346-1983, tersoff-465-1984, tersoff-275-1986, wachutka-8512-1986, inglesfield-57-1987, chen-923-1989, garcia-moliner-1405-1990, stampfl-8461-1992, garcia-moliner-332-1994, hummel-1620-1998, mahboob-201307r-2004, demkov-195306-2005, schleife-012014-2009, james-155439-2010, bravi-155445-2014, bk:inglesfield-2015}, %
the construction of Wannier functions \cite{kohn-809-1959, des-cloizeaux-a685-1964, des-cloizeaux-a698-1964, krieger-776-1967, kohn-2485-1973, rehr-1981-1974, rehr-448-1974, he-5341-2001}, %
impurities \cite{schmidt-425-1957, kohn-2485-1973, rehr-1981-1974, inkson-369-1980, blow-359-1980, burt-1825-1980, bylander-4157-1980, blow-3711-1982, blow-5267-1983, kostyrko-3241-1999, hjort-5245-2000, kostyrko-2458-2000, dwivedi-134304-2016}, %
high-energy electron diffraction \cite{buxton-3941-1977}, %
superlattices \cite{inglesfield-162-1971, schulman-4149-1981, krishnamurthy-1027-1985, mailhiot-8360-1986, ram-mohan-6151-1988, chen-923-1989, ghahramani-1102-1989, garcia-moliner-1405-1990, smith-173-1990, ferreira-8198-1998, kostyrko-2458-2000, reynoso-035301-2014}, %
heterostructures \cite{kurtin-756-1970, kurtin-3368-1971, osbourn-2124-1979, schulman-2346-1983, tersoff-4874-1984, marsh-285-1986, ko-9945-1988, ting-3583-1992, boykin-7670-1996, monch-5076-1996, ting-985-1999, demkov-195306-2005, monch-2-2014}, %
quantum wells \cite{brand-607-1987, chen-923-1989, garcia-moliner-1405-1990, boykin-8107-1996, boykin-7670-1996, ogawa-1527-1998}, %
magnetic systems \cite{maclaren-5470-1999, sanvito-11936-1999, mavropoulos-1088-2000, butler-054416-2001, dederichs-108-2002, velev-216601-2005, eames-252511-2006, lukashev-224414-2012, zhang-222401-2012}, %
electron transport \cite{ando-8017-1991, bowen-2754-1995, boykin-7670-1996, kemp-8349-1996, choi-2267-1999, kostyrko-3241-1999, maclaren-5470-1999, kostyrko-2458-2000, wortmann-165103-2002, tomfohr-245105-2002, tomfohr-59-2002, tomfohr-235105-2002, kostyrko-4393-2002, krstic-205319-2002, picaud-3731-2003, tomfohr-1542-2004, fagas-268-2004, khomyakov-195402-2004, pecchia-1497-2004, wang-016401-2004, pomorski-115408-2004, inglesfield-155120-2005, khomyakov-035450-2005, li-194113-2006, lee-215204-2007, zhang-035108-2007, rungger-035407-2008, sorensen-155301-2008, prodan-035124-2009, sorensen-205322-2009, varga-085102-2009, james-155439-2010, vergniory-544-2010, guan-1296-2011, jiang-057202-2012}, %
nanomaterials \cite{choi-2267-1999, kostyrko-3241-1999, ferreira-16040-2000, kostyrko-2458-2000, pomorski-115408-2004, xia-1597-2004, hod-114704-2006, hod-233401-2007, reuter-085412-2011, reuter-084707-2013, reynoso-035301-2014, szczesniak-355301-2016}, %
topological materials \cite{hatsugai-11851-1993, hatsugai-3697-1993, mong-125109-2011, avila-137-2012, dang-155307-2014, reynoso-035301-2014, tauber-115008-2015, dwivedi-134304-2016}, %
solar cells \cite{sulzer-3074-2016}, %
quantum size effects \cite{kalkstein-85-1971, cottey-1734-1971, cottey-2591-1972, cottey-2446-1973, brasher-4868-1980, inglesfield-57-1987, hod-114704-2006, hod-233401-2007, reuter-034703-2010, reuter-085412-2011, reuter-084707-2013}, %
and many others. However, these numerous applications often independently redevelop the main concepts and results of CBS, frequently with different notations or nomenclature. As a result, there are several formulations of CBS that appear unrelated at a quick inspection. Concrete examples of these formulations include %
the transfer or companion matrices \cite{schmidt-425-1957, cottey-1235-1971, wood-1400-1973, lee-4988-1981, lee-4997-1981, lee-355-1981, chang-3975-1982, schulman-2346-1983, strandberg-60-1983, biczo-51-1985, biczo-1992-1985, gies-267-1987, ram-mohan-6151-1988, ko-9945-1988, ghahramani-1102-1989, stampfl-8461-1992, bowen-2754-1995, boykin-8107-1996, ting-985-1999, wortmann-165103-2002, zhang-035108-2007, avila-137-2012, sanchez-soto-191-2012, reynoso-035301-2014, dwivedi-134304-2016}, 
the propagation matrix \cite{marcus-925-1968, jepsen-416-1971, bross-173-1977}, %
bivariational methods \cite{wachutka-3083-1982}, %
wavefunction matching techniques \cite{goodwin-205-1939, shockley-317-1939, marsh-285-1986, brand-607-1987, wachutka-8512-1986, inglesfield-57-1987, hummel-1620-1998, prodan-035128-2006}, %
and Green function-based approaches \cite{allen-917-1979, dy-4237-1979, mostoller-552-1979, brasher-4868-1980, kambe-443-1982, garcia-moliner-1405-1990, garcia-moliner-332-1994, umerski-5266-1997, reuter-085412-2011}. %
Note that these terms are very broad; \textit{e.g.},\ several different ``transfer matrices'' have been reported. To exacerbate these inconsistencies, many of the CBS derivations are unnecessarily tied to the choice of basis set and/or are laden with application-specific details. Consequently, the handful of studies that compare different CBS formulations \cite{hoffstein-99-1970, chang-3975-1982, wachutka-3083-1982, biczo-51-1985, wachutka-8512-1986, gies-267-1987, chen-923-1989, schulman-6282-1992, garcia-moliner-1405-1990, hummel-1620-1998, wortmann-165103-2002, pecchia-1497-2004, khomyakov-035450-2005} laid crucial groundwork but have yet to illuminate the full generality of CBS.

In this work we present (section \ref{sec:unified}) a unified perspective of CBS that delineates the key underpinnings of CBS that are often unstated or unclear. From these we develop an interpretation for CBS that relates to the amount of information in the material's Hamiltonian. We then use this realization to review previous studies (sections \ref{sec:cbs-wf}--\ref{sec:extended-coupling}) and demonstrate the commonalities among the various CBS formulations.

\section{Fundamentals and a Motivating Example}
\label{sec:example}
Similar to conventional band structure, the dispersion relation, $E(\vec{k})$, is a central quantity in CBS. The difference is that we now consider evanescent states in the material by letting the components of $\vec{k}$ be complex. As in conventional band structure, each $\vec{k}$ produces an eigenvalue problem that can be solved for $E$; in this way we can regard $E(\vec{k})$ as a multi-valued complex function \cite{bk:needham-1997}. Even though there will be multiple (and generally complex) $E$ associated with each complex $\vec{k}$, we are only interested in $\vec{k}$ that produce real $E$. Unless explicitly noted, we assume $E$ is real in what follows. This section uses complex analysis \cite{bk:needham-1997} to establish general properties of $E(\vec{k})$; methods for calculating $E(\vec{k})$ will be discussed in later sections. Without loss of generality [see point \ref{properties:general:reciprocal} below], we restrict $\mathrm{Re}(\vec{k})$ to the first Brillouin zone.

\begin{figure*}
\resizebox{6.5in}{!}{\includegraphics{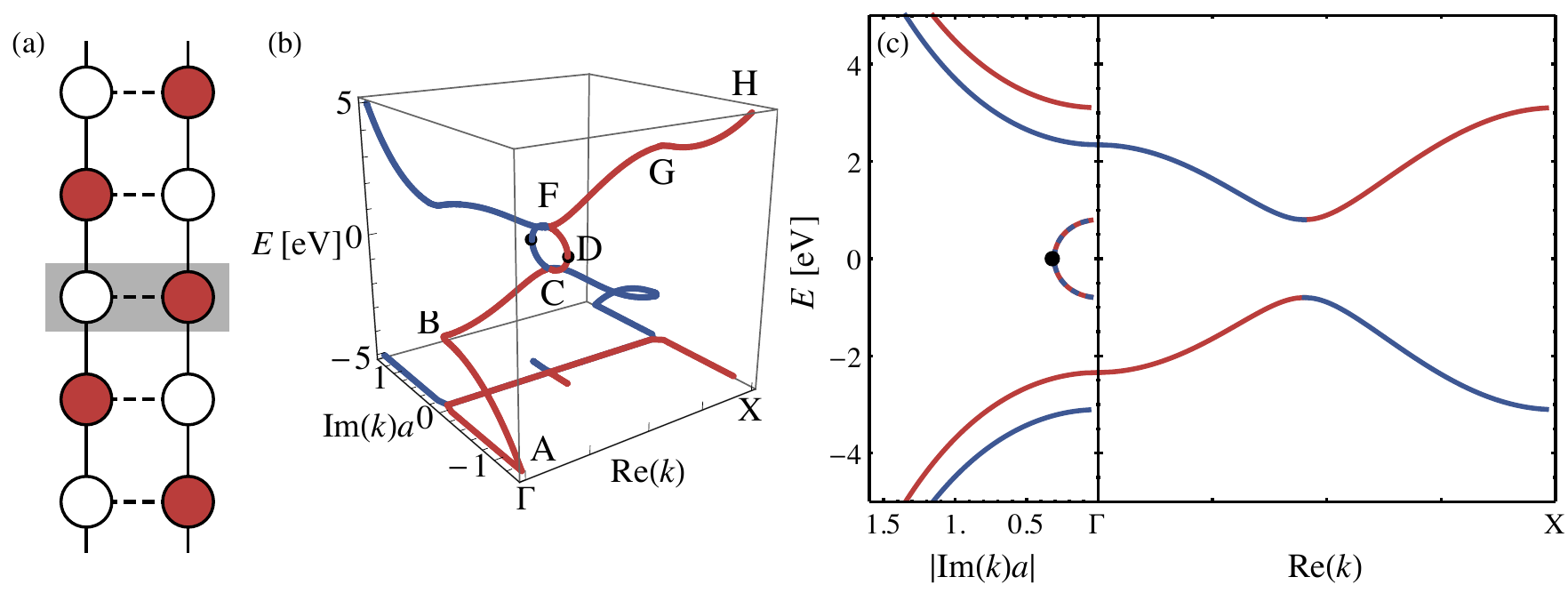}}
\caption{\label{fig:symmetric}Complex band structure, $E(\vec{k})$, for the one-dimensional, tight-binding model discussed in the text ($\alpha=0.8$ eV, $\beta_1=-0.4$ eV, $\beta_2=-1.3$ eV, and $a$ is the lattice constant). (a) A schematic of the system. The gray shading displays our choice of layer. (b) A three-dimensional view of the dispersion relation, illuminating Heine's ``lines of real $E$'' \cite{heine-300-1963, heine-1-1964} in complex $\vec{k}$ space. The two lines (red and blue) are also projected onto the base of the plot---the complex $\vec{k}$ plane---for additional perspective. The letters A--H trace out one of the ``lines of real energy'' and are references for the main discussion. (c) A standard, two-dimensional view of the dispersion relation. On the right is the conventional band structure, where $\vec{k}$ is real and in the first Brillouin zone. If a complex $\vec{k}$ produces a real energy $E$, the imaginary part of $\vec{k}$ (in magnitude) is plotted on the left. Black dots in (b) and (c) denote branch points of $E(\vec{k})$.}
\end{figure*}

As a concrete example, we begin by introducing a model system that exemplifies the properties of $E(\vec{k})$ we seek to discuss. This one-dimensional, two-band, tight-binding model is schematically depicted in Figure \ref{fig:symmetric}(a), with a layer shaded in gray. Each layer of the material has two orbitals, one with energy $\alpha$ (red circle) and the other with energy $-\alpha$ (white circle). Orbitals within the same layer are coupled by $\beta_1$ (dashed lines), and an orbital couples to the opposite-energy orbital in the two adjacent layers with $\beta_2$ (solid lines). The Hamiltonian is
\begin{equation}
\mathbf{H} = \sum_n \left[ \alpha \mathbf{a}_{n,+}^\dagger \mathbf{a}_{n,+} - \alpha \mathbf{a}_{n,-}^\dagger \mathbf{a}_{n,-} + \left( \beta_1 \mathbf{a}_{n,\pm}^\dagger \mathbf{a}_{n,\mp} + \beta_2 \mathbf{a}_{n+1,\pm}^\dagger \mathbf{a}_{n,\mp} + \mathrm{H.c.}\ \right) \right],
\label{eq:model-h}
\end{equation}
where $\mathbf{a}_{n,\pm}^{(\dagger)}$ is the annihilation (creation) operator for an electron in layer $n$'s orbital with energy $\pm\alpha$, and ``$\mathrm{H.c.}$''\ denotes Hermitian conjugate. Figure \ref{fig:symmetric}(b,c) displays $E(\vec{k})$, that is, the CBS, for this model system with $\alpha=0.8$ eV, $\beta_1=-0.4$ eV, and $\beta_2=-1.3$ eV; the lattice constant is taken to be $a=1$ (in arbitrary length units). These parameters are arbitrarily chosen to provide a representative example that clearly displays the key features of CBS. Examples of more realistic materials can be found in the literature \cite{chang-605-1982, dy-633-1982}.

Development of CBS began by considering the fundamental properties of $E(\vec{k})$ when $\vec{k}$ is only complex in one dimension \cite{kohn-809-1959, heine-300-1963, heine-1-1964, krieger-776-1967, prodan-035128-2006}. Extensions to higher-dimensional cases exist \cite{goodwin-205-1939, blount-305-1962, heine-300-1963, des-cloizeaux-a685-1964, des-cloizeaux-a698-1964, avron-85-1978, inkson-369-1980, avila-137-2012, szczesniak-355301-2016} but are uncommon and relatively undeveloped. Accordingly, we restrict our discussion to one-dimensional cases. This may correspond to an effectively one-dimensional material (\textit{e.g.},\ a nanotube or nanoribbon) or to a three-dimensional material with a surface, where translational symmetry is only broken in one direction. In the latter case, components of $\vec{k}$ parallel to the surface remain good quantum numbers; they are real and can be treated parametrically \cite{heine-300-1963, jones-443-1966, krieger-776-1967, kalkstein-85-1971}. Because our discussion thus focuses on one-dimensional $\vec{k}$, the vector notation will be suppressed.

$E(k)$ is generally complex for an arbitrary, complex $k$, and we are physically interested in real $E$. This is trivially satisfied when $k$ is real (\textit{i.e.},\ conventional band structure); the complex $k$ that lead to real $E$ constitute the CBS. Heine \cite{heine-300-1963, heine-1-1964} and Krieger \cite{krieger-776-1967} showed that these $k$ lie on ``lines of real energy'' in complex $k$-space, which (for our running example) are displayed in Figure \ref{fig:symmetric}(b). The red and blue lines in the figure are the two such lines in our model.

We see that the lines of real energy are continuous and that they only cross a particular energy once such that $E$ can be regarded as a parameter. Let us trace out the trajectory of one of these lines in Figure \ref{fig:symmetric}(b) (the other line is similar). When $E$ is very small (point A), the line is far from the real axis, but approaches the real axis as $E$ increases (toward B). In our example, $\mathrm{Re}(k)$ is constant during this process. Eventually, $E$ hits the edge of a conventional band (B), and the line turns $90^\circ$ to run along the real axis. This is a general result that we discuss in more detail below [point \ref{properties:realE:bandedge}]. $k$ runs along the real axis (B$\to$C) while $E$ continues to increase---forming the conventional band---until $E$ reaches the upper band edge (C). Similar to before, the line of real energy turns $90^\circ$ here and moves off into the complex plane. At some $E$ in the band gap (D) the line turns around and starts heading back toward the real axis. It makes another $90^\circ$ turn at the lower edge (F) of the next band and continues along the real $k$-axis to form the band (F$\to$G). Finally, at the upper edge of the band (G), the line of real energy makes one last $90^\circ$ turn into the complex plane. Further increases in $E$ take the line away from the real axis (toward H). For a slightly different perspective, Figure \ref{fig:symmetric}(c) also plots these lines of real energy in a more traditional style. The right-hand side displays the conventional band structure and the left-hand side shows $|\mathrm{Im}(k)|$ when $k$ is not real.

With this example in mind, let us now briefly discuss some general properties of $E(k)$. A more detailed discussion of these (and other) results is deferred to sections \ref{sec:properties} and \ref{sec:interpretations}. Of greatest importance, Kohn \cite{kohn-809-1959}, Blount \cite{blount-305-1962}, and Krieger \cite{krieger-776-1967} found that $E(k)$ is analytic everywhere except at isolated branch points (see Ref.\ \onlinecite{bk:needham-1997} for a review of branches and branch points). Recalling that $E(k)$ is a multi-valued function, each branch is essentially a band, where a branch point is the boundary between two bands. For reference, there are two branch points in our example that occur at conjugate $k$ values and are marked by black dots in Figure \ref{fig:symmetric}(b,c). It is generally the case that branch points occur in band gaps; that is, they are not on the real axis [see point \ref{properties:general:bp}]. This makes sense: The divide between two bands isn't in either band.

\begin{figure*}
\resizebox{6.5in}{!}{\includegraphics{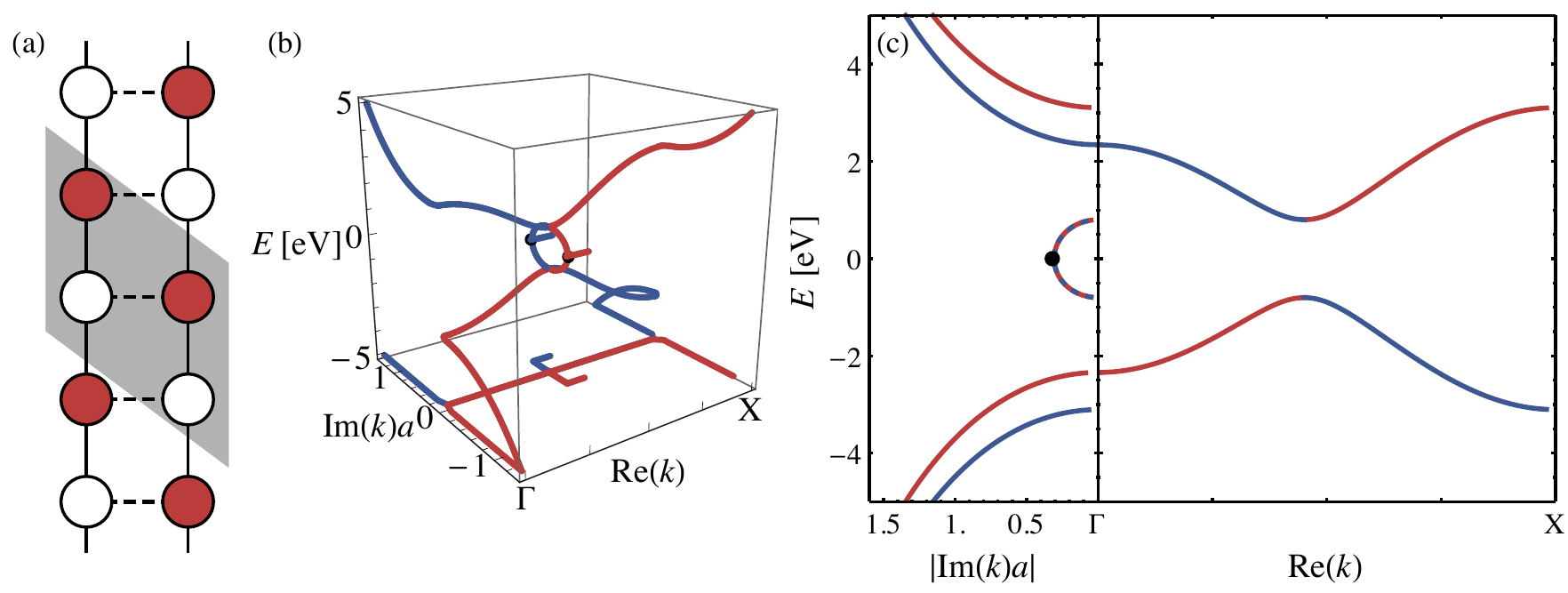}}
\caption{\label{fig:asymmetric}Complex band structure for the same model material as in Figure \ref{fig:symmetric}, but with a different choice of layers that lacks a mirror plane between layers. (a) Schematic of the model material with a layer highlighted in gray. (b) Three-dimensional view of the material's CBS with the new layers. The ``lines of real energy'' are largely indistinguishable from those in Figure \ref{fig:symmetric}(b). The only differences are around the branch points, where the lines now go around the branch points (and corresponding branch cuts) by making jaunts parallel (seemingly) to the real axis. (c) The corresponding two-dimensional view of the material's CBS. Because $\mathrm{Re}(k)$ is not plotted when $\mathrm{Im}(k)\neq0$, there is no apparent difference between this panel and Figure \ref{fig:symmetric}(c). Note that the lattice constant is different with this choice of layer (it is effectively double that from Figure \ref{fig:symmetric}). For ease of comparison, values of $k$ reported here have been shifted and scaled to use the same lattice constant as in Figure \ref{fig:symmetric}.}
\end{figure*}

In our example, the branch points were also the ``turn-around points'' for lines of real energy, where the lines reached their maximum distance from the real axis (within the band gap). This is not generally the case \cite{heine-300-1963}, but occurs when there is a mirror plane between the layers [see point \ref{properties:realE:branchpoint} below]. To emphasize this point, Figure \ref{fig:asymmetric} displays the CBS for the same model material with a different choice of ``layer''. Whereas there were only two orbitals in each layer before, there are now four [see Figure \ref{fig:asymmetric}(a)], and the mirror plane between layers is absent. A quick inspection of the ``two-dimensional view'' of CBS in Figure \ref{fig:asymmetric}(c) reveals essentially no differences from Figure \ref{fig:symmetric}(c), reinforcing the idea that the dispersion relation is a material property and independent of the arbitrary choice of layer. Examining the ``three-dimensional view'' [Figure \ref{fig:asymmetric}(b)] shows only one notable difference. The lines of real energy jut around the branch points (and corresponding branch cuts) rather than go through them. Accordingly, the lines of real energy in Figures \ref{fig:symmetric}(b) and \ref{fig:asymmetric}(b) differ only around the branch point and only in $\mathrm{Re}(k)$, which is relatively insignificant to $\mathrm{Im}(k)$ when $\mathrm{Im}(k)\neq0$. As we will see in later sections, even though the dispersion relation (\textit{i.e.},\ the set of eigenvalues from the time-independent Schr\"odinger equation) is largely independent of the choice of layer, the Bloch wavefunctions (the corresponding eigenfunctions) are much more sensitive to this choice.

\section{Unified Complex Band Structure}
\label{sec:unified}

Now that we have some familiarity with the fundamentals of CBS, we turn our focus to computational techniques and discuss methods for calculating CBS. Some general considerations can be found in Refs.\ \onlinecite{jones-443-1966, krieger-776-1967, chang-605-1982, bross-215-1989}, which review the fundamental properties of CBS as well. More specific (and direct) computational procedures have also been developed, as mentioned in section \ref{sec:introduction:background}, but almost always unnecessarily combine formulation with the model or basis set. This, in turn, obscures the inherent generality of CBS and has, perhaps, limited its utility. The goal of this section is to abstract the physical foundations of CBS from the existing literature and to demonstrate that this underlying structure \textit{implies} the existence of CBS. As in section \ref{sec:example}, we focus our attention on one-dimensional CBS.

The key premise of CBS is repetition of a material's ``repeat unit''. The repeat unit may be infinitely tessellated (becoming the ``unit cell''), which results in translational symmetry and conventional band structure. But it may not be. Disorder, perhaps caused by a surface/interface or an impurity, disrupts the repetition. The common factor in all of these cases is that the repeat unit is prevalent throughout the system; that is, \textit{most} layers are identical to the repeat unit.

\subsection{Mathematics of Repetition}
\label{sec:unified:repeat}
As seen in our examples from section \ref{sec:example}, the first step is to identify each ``layer'' within the system. The layer doesn't necessarily have to be an atomic layer, but is simply a way to partition the system and define the repeat units. Mathematically, this is accomplished using orthogonal (Hermitian) projection operators \cite{bk:yanai-2011-ch2}. For example, let $\mathbf{P}_j$ be the projector for layer $j$. Then, $\mathbf{H}_{j,j} \equiv \mathbf{P}_j \mathbf{H} \mathbf{P}_j$ denotes the ``block'' of the Hamiltonian corresponding to layer $j$; likewise,
\begin{equation}
\mathbf{H}_{j,k} \equiv \mathbf{P}_j \mathbf{H} \mathbf{P}_k
\label{eq:def-block}
\end{equation}
is the coupling between layers $j$ and $k$. Because it is built with operators, this projector-based notion of a ``matrix block'' sidesteps any concerns about basis set locality or nonorthogonality and permits a discussion of CBS that is unencumbered by the specific choice of basis set \cite{strandberg-60-1983}. Briefly returning to our examples in section \ref{sec:example}, Figure \ref{fig:symmetric} specified the projectors
\[
\mathbf{P}_n = \mathbf{a}_{n,+}^\dagger \mathbf{a}_{n,+} + \mathbf{a}_{n,-}^\dagger \mathbf{a}_{n,-},
\]
whereas the larger layers in Figure \ref{fig:asymmetric} used
\[
\mathbf{P}_n = \mathbf{a}_{2n+1,+}^\dagger \mathbf{a}_{2n+1,+} + \mathbf{a}_{2n,-}^\dagger \mathbf{a}_{2n,-} + \mathbf{a}_{2n,+}^\dagger \mathbf{a}_{2n,+} + \mathbf{a}_{2n-1,-}^\dagger \mathbf{a}_{2n-1,-}.
\]

We also require (for now) the projectors to be chosen such that each layer only couples to its nearest neighbors. The size of each layer can always be made larger to satisfy this condition, and we will discuss relaxing it in section \ref{sec:extended-coupling}. Mathematically, the Hamiltonian then assumes a block tridiagonal form,
\begin{equation}
\mathbf{H} = \left[
\begin{array}{ccccc}
\mathbf{H}_{1,1} & \mathbf{H}_{1,2} & \mathbf{0} & \mathbf{0} & \cdots \\
\mathbf{H}_{2,1} & \mathbf{H}_{2,2} & \mathbf{H}_{2,3} & \mathbf{0} & \cdots \\
\mathbf{0} & \mathbf{H}_{3,2} & \mathbf{H}_{3,3} & \mathbf{H}_{3,4} & \cdots \\
\mathbf{0} & \mathbf{0} & \mathbf{H}_{4,3} & \mathbf{H}_{4,4} & \cdots \\
\vdots & \vdots & \vdots & \vdots & \ddots \\
\end{array}
\right],
\label{eq:bt-h}
\end{equation}
with $\mathbf{H}_{j,k}=\mathbf{H}_{k,j}^\dagger$ (by Hermiticity). As enumerated in Eq.\ \eqref{eq:bt-h}, the system has a ``surface'' to the left of layer 1 (there is nothing to the left of layer 1). Accordingly, layer 1 is the surface layer, layer 2 the first subsurface layer, and so forth to the ``bulk limit'' at $\infty$ \cite{reuter-034703-2010, reuter-085412-2011, reuter-084707-2013}. In the absence of a surface, the indices would extend to $-\infty$.

If we further neglect other forms of disorder (such as surface reconstructions or impurities), each layer becomes identical to the others. This is the repeat unit. The Hamiltonian is thus also block Toeplitz; that is, blocks are the same along a particular diagonal. We denote $\mathbf{H}_{j,j}=\mathbf{H}_\mathrm{D}$ and $\mathbf{H}_{j,j-1}=\mathbf{H}_\mathrm{S}$ ($\mathbf{H}_{j-1,j}=\mathbf{H}_\mathrm{S}^\dagger$) such that
\begin{equation}
\mathbf{H} = \left[
\begin{array}{ccccc}
\mathbf{H}_\mathrm{D} & \mathbf{H}_\mathrm{S}^\dagger & \mathbf{0} & \mathbf{0} & \cdots \\
\mathbf{H}_\mathrm{S} & \mathbf{H}_\mathrm{D} & \mathbf{H}_\mathrm{S}^\dagger & \mathbf{0} & \cdots \\
\mathbf{0} & \mathbf{H}_\mathrm{S} & \mathbf{H}_\mathrm{D} & \mathbf{H}_\mathrm{S}^\dagger & \cdots \\
\mathbf{0} & \mathbf{0} & \mathbf{H}_\mathrm{S} & \mathbf{H}_\mathrm{D} & \cdots \\
\vdots & \vdots & \vdots & \vdots & \ddots \\
\end{array}
\right].
\label{eq:btbt-h}
\end{equation}
(The ``$\mathrm{D}$'' and ``$\mathrm{S}$'' subscripts stand for ``diagonal'' and ``sub-/super-diagonal'', respectively.) For the time being, we assume that $\mathbf{H}_\mathrm{S}$ is invertible. Section \ref{sec:singular} will discuss relaxing this condition.

For computational purposes (\textit{e.g.},\ the examples in section \ref{sec:example}), we assume that each layer can be sufficiently described by $M$ basis functions, such that the operators $\mathbf{H}_\mathrm{D}$ and $\mathbf{H}_\mathrm{S}$ can be represented by $M\times M$ matrices. Matrix representations for the small layers in Figure \ref{fig:symmetric} are $2\times2$,
\[
\mathbf{H}_\mathrm{D} = \left[ \begin{array}{cc} \alpha & \beta_1^\ast \\ \beta_1 & -\alpha \end{array} \right] \text{ and } \mathbf{H}_\mathrm{S} = \left[ \begin{array}{cc} 0 & \beta_2 \\ \beta_2 & 0 \end{array} \right],
\]
whereas the larger layers in Figure \ref{fig:asymmetric} require $4\times4$ matrices,
\[
\mathbf{H}_\mathrm{D} = \left[ \begin{array}{cccc} \alpha & 0 & \beta_2^\ast & 0 \\ 0 & \alpha & \beta_1^\ast & \beta_2^\ast \\ \beta_2 & \beta_1 & -\alpha & 0 \\ 0 & \beta_2 & 0 & -\alpha \end{array} \right] \text{ and } \mathbf{H}_\mathrm{S} = \left[ \begin{array}{cccc} 0 & 0 & \beta_2 & \beta_1 \\ 0 & 0 & 0 & \beta_2 \\ 0 & 0 & 0 & 0 \\ 0 & 0 & 0 & 0 \end{array} \right].
\]
Note that $\mathbf{H}_\mathrm{S}$ for the larger layers is not invertible, and provides an example for the discussion in section \ref{sec:singular}. These basis set and matrix details are only necessary for computation and (unless otherwise noted) will not be considered when formulating or discussing the theory of CBS.

\subsection{Block Tridiagonal, Block Toeplitz Matrices}
\label{sec:unified:btbtm}
This block tridiagonal and block Toeplitz structure in the Hamiltonian is foundational to CBS, as it mathematically leads to CBS. We now detail this idea.

Most applications aim to either diagonalize or (essentially) invert the Hamiltonian. The former appears in the context of the time-independent Schr\"odinger equation,
\begin{equation}
\mathbf{H} \ket{\psi} = E \ket{\psi},
\label{eq:schrodinger}
\end{equation}
and the latter in finding the (retarded/advanced) Green function (GF) \cite{bk:economou-2006},
\begin{equation}
\mathbf{G}(E) = \lim_{\eta \to 0^\pm} \left[ (E+ i \eta)\mathbf{I} - \mathbf{H} \right]^{-1}.
\label{eq:gf}
\end{equation}
As foreshadowed, the block tridiagonal and block Toeplitz structure of $\mathbf{H}$ greatly aids in these processes.

To begin, consider the amount of information in $\mathbf{H}$. All that is needed to \textit{completely} describe our system is $\mathbf{H}_\mathrm{D}$, $\mathbf{H}_\mathrm{S}$, and the number of layers (\textit{i.e.},\ the number of block rows/columns). With respect to the size of our system (the number of layers) this is $\mathcal{O}(1)$ information; that is, the amount of information \textit{does not change} with system size. Because $\mathbf{H}$ uniquely determines the eigenvalues (eigenvectors) and the GF, there must only be $\mathcal{O}(1)$ information in these quantities. CBS is this $\mathcal{O}(1)$ information. All reported formulations of CBS access this information to solve specific problems, \textit{e.g.},\ diagonalizing $\mathbf{H}$ for the time-independent Schr\"odinger equation. In this manner, CBS describes, and can be used to understand, all of a material's static and dynamic electronic properties.

Such an analysis of a matrix's information content is not new to the mathematics community, and is one facet of studying (block) quasi-separable (sometimes called rank-structured) matrices \cite{bk:vandebril-2008-v1, bk:vandebril-2008-v2}. As it suffices for our purposes (a more rigorous definition is presented in %
\footnote{A rigorous definition of a block quasi-separable matrix can be found on p. 482 of Ref.\ \onlinecite{bk:vandebril-2008-v1}. Suppose $\mathbf{A}$ is a block matrix with $N$ block rows and $N$ block columns. If $\mathbf{A}$ is block quasi-separable, then its blocks can be written as
\[
\mathbf{A}_{j,k} = \begin{cases}
\mathbf{U}_j \mathbf{T}_{j-1} \mathbf{T}_{j-2} \cdots \mathbf{T}_{k+1} \mathbf{V}_k & \mbox{if } 1\le k < j \le N \\
\mathbf{D}_j & \mbox{if } 1\le j = k \le N \\
\mathbf{P}_j \mathbf{R}_{j+1} \mathbf{R}_{j+2} \cdots \mathbf{R}_{k-1} \mathbf{Q}_k & \mbox{if } 1\le j < k \le N \\
\end{cases},
\]
where the $\mathbf{U}_j$, $\mathbf{T}_j$, $\mathbf{V}_j$, $\mathbf{D}_j$, $\mathbf{P}_j$, $\mathbf{R}_j$, and $\mathbf{Q}_j$ sequences are a set of generators for $\mathbf{A}$.
}%
), a block quasi-separable matrix has certain sub-parts with specific rank. For the present discussion, the sub-parts that contain a sub-diagonal (or super-diagonal) block and the bottom-left (top-right) corner, but do not contain a diagonal block, will have rank equal to that of $\mathbf{H}_\mathrm{S}$. Graphical examples of these sub-parts are
\[
\left[ \begin{array}{cccc}
\mathbf{H}_\mathrm{D} & \tikzmarkin[hor=style grey]{e1} \mathbf{H}_\mathrm{S}^\dagger & \mathbf{0} & \mathbf{0} \tikzmarkend{e1} \\
\mathbf{H}_\mathrm{S} & \mathbf{H}_\mathrm{D} & \mathbf{H}_\mathrm{S}^\dagger & \mathbf{0} \\
\mathbf{0} & \mathbf{H}_\mathrm{S} & \mathbf{H}_\mathrm{D} & \mathbf{H}_\mathrm{S}^\dagger \\
\mathbf{0} & \mathbf{0} & \mathbf{H}_\mathrm{S} & \mathbf{H}_\mathrm{D} \\
\end{array} \right]
\text{ and }
\left[ \begin{array}{cccc}
\mathbf{H}_\mathrm{D} & \mathbf{H}_\mathrm{S}^\dagger & \mathbf{0} & \mathbf{0} \\
\mathbf{H}_\mathrm{S} & \mathbf{H}_\mathrm{D} & \mathbf{H}_\mathrm{S}^\dagger & \mathbf{0} \\
\tikzmarkin[hor=style grey]{e2} \mathbf{0} & \mathbf{H}_\mathrm{S} & \mathbf{H}_\mathrm{D} & \mathbf{H}_\mathrm{S}^\dagger \\
\mathbf{0} & \mathbf{0} \tikzmarkend{e2} & \mathbf{H}_\mathrm{S} & \mathbf{H}_\mathrm{D} \\
\end{array} \right].
\]
Clearly, block tridiagonal matrices (with or without the block Toeplitz structure) belong to the class of block quasi-separable matrices.

Ramifications of this block quasi-separable structure have been discussed in numerous contexts \cite{bk:vandebril-2008-v1, bk:vandebril-2008-v2}. In many cases, they include analytical characterizations and/or fast numerical algorithms for inverting or diagonalizing a matrix. References \onlinecite{mostoller-552-1979, mostoller-6168-1982, rozsa-447-1989, godfrin-7843-1991, meurant-707-1992, huang-7919-1997, tomfohr-1542-2004, techrpt:jain-2007, koulaei-223-2007, cauley-043713-2011, reuter-014009-2012, boffi-015001-2015} overview specific results for block tridiagonal matrices. In the end, these results stem from finding and exploiting the ``generators'' of the matrix \cite{eidelman-187-2005}; that is, the minimal information needed to generate the entire matrix, its spectral decomposition, or its inverse. The generators are not unique, and some generators may be preferential to others for specific operations (such as inverting or diagonalizing the matrix).

Returning to our block tridiagonal and block Toeplitz Hamiltonian, we see that the set $\{ \mathbf{H}_\mathrm{D}, \mathbf{H}_\mathrm{S}, N\}$ is a generator for $\mathbf{H}$ (where $N$ is the number of layers). We will show in the following sections that the various formulations of CBS produce different generators, as needed for specific applications, but that they all represent the same information (\textit{i.e.},\ CBS). For example, the transfer matrix \cite{lee-355-1981} and the companion matrix \cite{chang-3975-1982} help in diagonalizing $\mathbf{H}$, whereas generators for (in effect) inverting $\mathbf{H}$ to obtain the GF have also been developed \cite{dy-4237-1979, umerski-5266-1997}. In the following sections we derive some of these generators (that is, formulate CBS) and show how they aid in diagonalizing or inverting $\mathbf{H}$. Along the way, we prove that these generators embody the same information (CBS) and then discuss some uses and applications of CBS.

\section{Complex Band Structure from Wavefunctions}
\label{sec:cbs-wf}

Our first formulation of CBS is a wavefunction-based approach that begins with the time-independent Schr\"odinger equation [Eq.\ \eqref{eq:schrodinger}]. In this sense, we start with a particular (real) energy and aim to find the real or complex values of $k$ that produce states with that energy. In the language of complex analysis, we are calculating the multi-valued function $k(E)$, which is essentially the ``inverse'' of the traditional dispersion relation, $E(k)$. In other words, what are the complex (possibly real) wavevectors that produce to states with a given real energy? As we now see, the transfer matrix \cite{lee-355-1981} provides one method for answering this question.

We start by applying a projector (for an arbitrary layer) to the time-independent Schr\"odinger equation. Writing $\ket{\psi_j}\equiv\mathbf{P}_j\ket{\psi}$ and specifically considering layer $n$,
\begin{align}
E \mathbf{P}_n \ket{\psi} & = \mathbf{P}_n \mathbf{H} \ket{\psi}, \nonumber \\
E \ket{\psi_n} & = \mathbf{H}_{n,n-1} \ket{\psi_{n-1}} + \mathbf{H}_{n,n} \ket{\psi_{n}} + \mathbf{H}_{n,n+1} \ket{\psi_{n+1}}, \nonumber \\
& = \mathbf{H}_\mathrm{S} \ket{\psi_{n-1}} + \mathbf{H}_\mathrm{D} \ket{\psi_{n}} + \mathbf{H}_\mathrm{S}^\dagger \ket{\psi_{n+1}},
\label{eq:schrodinger-recurrence}
\end{align}
where we have utilized both the block tridiagonal and block Toeplitz structure of $\mathbf{H}$. If we now define a ``supercell'' wavefunction,
\begin{equation}
\ket{\Psi_{n+1}} \equiv \left[ \begin{array}{c} \ket{\psi_{n+1}} \\ \ket{\psi_n} \end{array} \right],
\label{eq:def-wf-supercell}
\end{equation}
Eq.\ \eqref{eq:schrodinger-recurrence} can be combined with the tautology $\ket{\psi_n}=\ket{\psi_n}$ to write
\begin{align}
\ket{\Psi_{n+1}} & = \left[ \begin{array}{c} -\mathbf{H}_\mathrm{S}^{-\dagger} \mathbf{H}_\mathrm{S} \ket{\psi_{n-1}} + \mathbf{H}_\mathrm{S}^{-\dagger} \left( E \mathbf{I} - \mathbf{H}_\mathrm{D} \right) \ket{\psi_n} \\ \ket{\psi_n} \end{array} \right] \nonumber \\
& = \left[ \begin{array}{cc} \mathbf{H}_\mathrm{S}^{-\dagger} \left( E \mathbf{I} - \mathbf{H}_\mathrm{D} \right) & -\mathbf{H}_\mathrm{S}^{-\dagger} \mathbf{H}_\mathrm{S} \\ \mathbf{I} & \mathbf{0} \end{array} \right] \left[ \begin{array}{c} \ket{\psi_n} \\ \ket{\psi_{n-1}} \end{array} \right] \nonumber \\
& = \mathbf{T} \ket{\Psi_{n}}, \label{eq:supercell-recurrence}
\end{align}
where $^{-\dagger}$ denotes the inverse Hermitian conjugate. In Eq.\ \eqref{eq:supercell-recurrence},
\begin{equation}
\mathbf{T} = \left[ \begin{array}{cc}
\mathbf{H}_\mathrm{S}^{-\dagger} \left( E \mathbf{I} - \mathbf{H}_\mathrm{D} \right) & -\mathbf{H}_\mathrm{S}^{-\dagger} \mathbf{H}_\mathrm{S} \\
\mathbf{I} & \mathbf{0}
\end{array} \right]
\label{eq:transfer-matrix}
\end{equation}
is the transfer matrix, which provides a convenient way to calculate the supercell wavefunction for any layer subject to some base case,
\[
\ket{\Psi_n} = \mathbf{T}^{n-1} \ket{\Psi_1}.
\]
Note that $\mathbf{T}$ depends on $E$ and that transfer matrices of this form have been discussed in more general contexts \cite{molinari-983-1997, molinari-8553-1998, molinari-4081-2003}.

The link between the transfer matrix and CBS results from an invocation of Bloch's theorem [Eq.\ \eqref{eq:bloch-theorem}]. Accordingly,
\[
\ket{\Psi_{n+1}} = \left[ \begin{array}{c} \ket{\psi_{n+1}} \\ \ket{\psi_n} \end{array} \right] = \left[ \begin{array}{c} e^{ika} \ket{\psi_n} \\ e^{ika} \ket{\psi_{n-1}} \end{array} \right] = e^{ika} \ket{\Psi_{n}}.
\]
Substituting this into Eq.\ \eqref{eq:supercell-recurrence} yields
\begin{equation}
e^{ika} \ket{\Psi_{n}} = \mathbf{T} \ket{\Psi_{n}};
\label{eq:transfer-eigval}
\end{equation}
that is, the eigenvalues of $\mathbf{T}$ are essentially the wavevectors that produce states with energy $E$. Consequently, the CBS for a given energy can be obtained from the spectrum of $\mathbf{T}$. As a side note, the CBS of our model system in Figure \ref{fig:symmetric} was obtained from the eigenvalues of $\mathbf{T}$ at each $E$.

\section{Complex Band Structure from Green Functions}
\label{sec:cbs-gf}
In the previous section we used wavefunction arguments to derive the transfer matrix and, ultimately, CBS. We now work through a GF-based approach to CBS based on matrix M\"obius transformations (MMTs) \cite{schwarz-1913-1981, umerski-5266-1997} and surface GFs. In essence, the inverse (\textit{i.e.},\ GF) of a block tridiagonal Hamiltonian can be completely obtained from surface GFs, and MMTs exploit the block Toeplitz structure to provide the surface GFs. This formulation combines and summarizes aspects of Refs.\ \onlinecite{umerski-5266-1997, godfrin-7843-1991, reuter-085412-2011, boffi-015001-2015}, and is a bit more complicated than the derivation of the transfer matrix. To aid in the discussion, we will present various densities of states (DOSs) for our model systems in section \ref{sec:example}. A DOS is related to a retarded GF by \cite{bk:economou-2006}
\begin{equation}
\rho(E) = -\frac{1}{\pi} \mathrm{Im}\left( \mathrm{Tr}\left[ \mathbf{G}(E) \right] \right),
\label{eq:dos}
\end{equation}
where ``$\mathrm{Tr}$'' is the trace.

Section \ref{sec:cbs-gf:mmts} first overviews some pertinent properties of MMTs. Then, section \ref{sec:cbs-gf:sgf} discusses the utility of MMTs in calculating surface GFs. Finally, section \ref{sec:cbs-gf:derive-cbs-gf} puts all of these elements together to arrive at CBS via the GF.

\subsection{Matrix M\"obius Transformations}
\label{sec:cbs-gf:mmts}
The MMT \cite{schwarz-1913-1981, umerski-5266-1997} is a generalization of the M\"obius (bilinear) transformation from complex variables \cite{bk:needham-1997} to matrices. Let $\mathbf{m}_{11}$, $\mathbf{m}_{12}$, $\mathbf{m}_{21}$, $\mathbf{m}_{22}$, and $\mathbf{z}$ be complex $n\times n$ matrices; a MMT is a mapping with the form
\begin{equation}
\mathbf{M} \bullet \mathbf{z} \equiv \left( \mathbf{m}_{11} \mathbf{z} + \mathbf{m}_{12} \right) \left( \mathbf{m}_{21} \mathbf{z} + \mathbf{m}_{22} \right)^{-1}.
\label{eq:matrix-mt}
\end{equation}
Similar to the complex M\"obius transformation \cite{bk:needham-1997}, we can associate with $\mathbf{M}$ the $2n\times 2n$ matrix \cite{umerski-5266-1997}
\[
\mathbf{M} = \left[ \begin{array}{cc} \mathbf{m}_{11} & \mathbf{m}_{12} \\ \mathbf{m}_{21} & \mathbf{m}_{22} \end{array} \right].
\]
A MMT and its matrix representation will be used interchangeably throughout this work. Note that the ``$\bullet$'' in Eq.\ \eqref{eq:matrix-mt} emphasizes that $\mathbf{M}\bullet\mathbf{z}$ is not a canonical matrix-matrix product. Lastly, it is easily verified that the application of MMTs is associative; that is, for any two MMTs $\mathbf{M}_1$ and $\mathbf{M}_2$,
\begin{equation}
\mathbf{M}_1 \bullet \left( \mathbf{M}_2 \bullet \mathbf{z} \right) = \left( \mathbf{M}_1 \mathbf{M}_2 \right) \bullet \mathbf{z},
\label{eq:mmt-associative}
\end{equation}
where $\mathbf{M}_1\mathbf{M}_2$ is the usual matrix-matrix product.

\subsection{Surface Green Functions}
\label{sec:cbs-gf:sgf}
As we will see in section \ref{sec:cbs-gf:derive-cbs-gf}, the GF for a block tridiagonal Hamiltonian can be formulated in terms of surface GFs. This subsection is, therefore, devoted to surface GFs, which have been thoroughly discussed elsewhere \cite{mostoller-552-1979, lopez-sancho-851-1985, umerski-5266-1997, velev-r637-2004, hod-114704-2006, rungger-035407-2008, reuter-085412-2011}. Herein we develop a layer-by-layer approach for surface GFs that parallels the transfer matrix for wavefunctions.

Before proceeding, we need to define the term ``surface GF'' and specify our notation. Using the layer enumeration in Eq.\ \eqref{eq:bt-h}, the surface GF is the $(1,1)$ block of the total system GF; that is,
\begin{equation}
\widetilde{\mathbf{G}}_\infty^\mathrm{R}(E) = \mathbf{P}_1 \mathbf{G}(E) \mathbf{P}_1 \equiv \mathbf{G}_{1,1}(E).
\label{eq:surface-gf}
\end{equation}
Three comments are needed to clarify our notation for surface GFs. First, the tilde signifies a surface GF, as opposed to the full system GF in Eq.\ \eqref{eq:gf}. Second, the surface GF depends on the orientation of the material relative to the surface. For example, the system described in Eq.\ \eqref{eq:bt-h} places the material to the right of the surface, where the surface is effectively between layers 0 and 1. We use the superscript ``$\mathrm{R}$'' to label this case; ``$\mathrm{L}$'' is used when the material is to the left of the surface. For reference, a left surface GF is the bottom-right block of the total system GF. Third, surface GFs for systems containing a finite number of layers will also be important; $\widetilde{\mathbf{G}}_N^\mathrm{L/R}(E)$ denotes the left/right surface GF for a system with $N$ layers. As demonstrated in Eq.\ \eqref{eq:surface-gf}, $\widetilde{\mathbf{G}}_\infty^\mathrm{L/R}(E)$ is the left/right surface GF for a semi-infinite system.

We now show that MMTs can build surface GFs in a layer-by-layer fashion. To overview the procedure, we use L\"owdin partitioning \cite{lowdin-12-1963} (which is essentially an embedding technique \cite{inglesfield-3795-1981, bk:inglesfield-2015}) to construct an effective Hamiltonian for the surface layer that (i) only exists in the surface layer and (ii) is rigorously equivalent to the total system Hamiltonian within the surface layer. This happens at the expense of having an energy-dependent effective Hamiltonian. Because the surface GF is the block of $\mathbf{G}$ for the surface layer, the surface GF is identical to the GF of this effective Hamiltonian. Equation \eqref{eq:sgf:layer-relate} formally states this result, and also relates the surface GF for a system with $N$ layers to the surface GF for a system with $N-1$ layers. MMTs then exploit Eq.\ \eqref{eq:sgf:layer-relate} to obtain the surface GF for an arbitrary number of layers; this realization leads to CBS in the next section. In what follows we detail this idea for right surface GFs, noting that a similar procedure is used for left surface GFs.

\begin{figure}
\resizebox{3.25in}{!}{\includegraphics{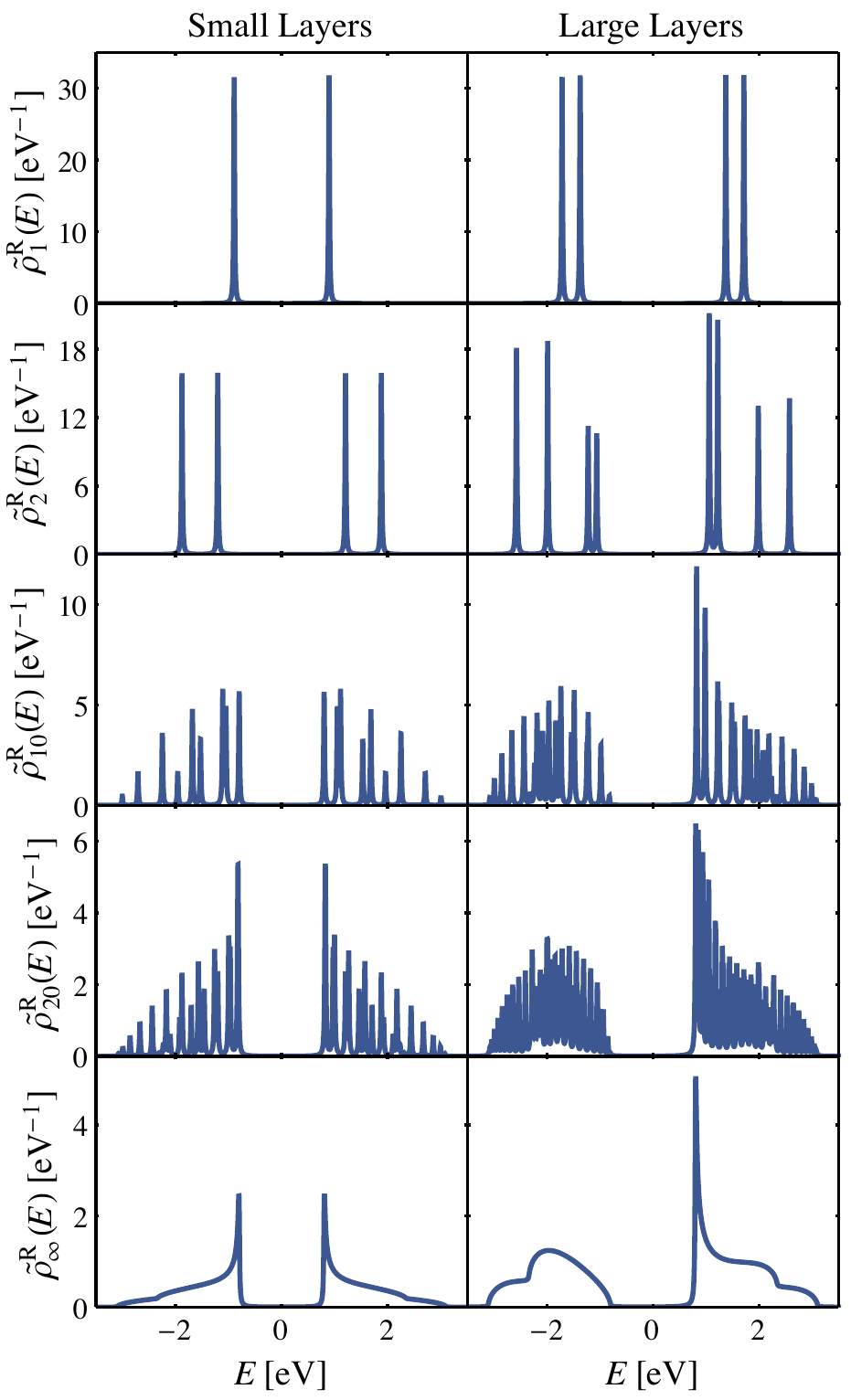}}
\caption{\label{fig:sgf}Right surface DOSs, given by Eq.\ \eqref{eq:sdos}, for the example systems discussed in section \ref{sec:example}. Left and right columns: Layers chosen using the projectors from Figures \ref{fig:symmetric} and \ref{fig:asymmetric}, respectively. Each row corresponds to systems with a different number of layers. From top to bottom: 1, 2, 10, 20, and the semi-infinite limit. With only 1 layer, the surface DOS shows isolated states (essentially $\delta$ functions). Bands grow as more layers are added, until the surface DOSs converge to the semi-infinite limit. An imaginary energy component of 10~meV was used [see Eq.\ \eqref{eq:gf}].}
\end{figure}

For concreteness, Figure \ref{fig:sgf} displays various surface DOSs,
\begin{equation}
\widetilde{\rho}_n^\mathrm{L/R}(E) = \frac{-1}{\pi} \mathrm{Im} \left( \mathrm{Tr}\left[ \widetilde{\mathbf{G}}_n^\mathrm{L/R}(E) \right] \right),
\label{eq:sdos}
\end{equation}
for our ongoing example (using the layers from both Figures \ref{fig:symmetric} and \ref{fig:asymmetric}). When there are only a few layers in the system, the surface DOS is essentially a collection of $\delta$ functions (one for each eigenvalue of $\mathbf{H}$). As the system gets larger, the $\delta$ functions coalesce into bands, ultimately giving rise to the continuous surface DOSs seen for the semi-infinite systems. When there is a mirror plane between the layers (as in Figure \ref{fig:symmetric}), the valence and conduction bands are anti-symmetrically shaped. There is no such similarity between the bands when the mirror plane is absent (\textit{e.g.},\ the layers of Figure \ref{fig:asymmetric}).

As in the derivation of the transfer matrix, we begin with the time-independent Schr\"odinger equation. To set up L\"owdin partitioning, suppose that $\mathbf{P}_1$ is the projector for our surface layer and $\mathbf{Q}_1\equiv\mathbf{I}-\mathbf{P}_1=\mathbf{P}_2+\mathbf{P}_3+\ldots$\ is the projector for the rest of the system. If we substitute $\mathbf{I}=\mathbf{P}_1+\mathbf{Q}_1$ before and after $\mathbf{H}$ in Eq.\ \eqref{eq:schrodinger}, distribute, and rearrange terms, we get the matrix equation
\begin{equation}
\left[ \begin{array}{cc} \mathbf{P}_1 \mathbf{HP}_1 & \mathbf{P}_1\mathbf{HQ}_1 \\ \mathbf{Q}_1\mathbf{HP}_1 & \mathbf{Q}_1\mathbf{HQ}_1 \end{array} \right] \left[ \begin{array}{c} \mathbf{P}_1 \ket{\psi} \\ \mathbf{Q}_1 \ket{\psi} \end{array} \right] = E \left[ \begin{array}{c} \mathbf{P}_1 \ket{\psi} \\ \mathbf{Q}_1 \ket{\psi} \end{array} \right].
\label{eq:sgf:lowdin}
\end{equation}
The bottom row can be rearranged to produce
\[
\left( E \mathbf{Q}_1 - \mathbf{Q}_1\mathbf{HQ}_1 \right) \mathbf{Q}_1 \ket{\psi} = \mathbf{Q}_1\mathbf{HP}_1 \ket{\psi}.
\]
We observe that the term in parentheses on the left-hand side is the inverse of the GF for the isolated system of all non-surface layers. If we invert this expression \footnote{Note that the inverse of $E\mathbf{Q}_1-\mathbf{Q}_1 \mathbf{HQ}_1$ is not necessarily well-defined because $\mathbf{Q}_1$ is not full rank. The final result, Eq.\ \eqref{eq:sgf:eff-h}, sandwiches this GF between $\mathbf{Q}_1$ projectors, essentially allowing us to use a pseudo-inverse \cite{bk:strang-2006} to mollify the problem.} to obtain the GF for that subsystem (denoted by $\mathbf{G}_{\mathbf{Q}_1}$), we get
\begin{equation}
\mathbf{Q}_1 \ket{\psi} = \mathbf{G}_{\mathbf{Q}_1}(E) \mathbf{Q}_1 \mathbf{HP}_1 \ket{\psi}.
\label{eq:sgf:pq-psi}
\end{equation}
This gives us the wavefunction outside the surface layer ($\mathbf{Q}_1\ket{\psi}$) in terms of the wavefunction in the surface layer ($\mathbf{P}_1\ket{\psi}\equiv\ket{\psi_1}$). Substituting Eq.\ \eqref{eq:sgf:pq-psi} into the top row of Eq.\ \eqref{eq:sgf:lowdin} and rearranging terms yields
\[
\mathbf{P}_1\mathbf{HP}_1 \ket{\psi_1} + \mathbf{P}_1 \mathbf{HQ}_1 \mathbf{G}_{\mathbf{Q}_1}(E) \mathbf{Q}_1 \mathbf{HP}_1 \ket{\psi_1} = E \ket{\psi_1}.
\]
We now have a time-independent Schr\"odinger equation exclusively in the surface layer, where the effective Hamiltonian is
\begin{equation}
\mathbf{P}_1 \mathbf{HP}_1 + \mathbf{P}_1\mathbf{HQ}_1 \mathbf{G}_{\mathbf{Q}_1}(E) \mathbf{Q}_1 \mathbf{HP}_1.
\label{eq:sgf:eff-h}
\end{equation}
Mathematically, this effective Hamiltonian is related to Schur complements \cite{bk:zhang-2005}. The second, energy-dependent term is sometimes called a self-energy, and can be interpreted as an embedding potential or as an open-system boundary condition.

The surface GF for our total system is then obtained by inverting this effective Hamiltonian within the surface layer's subspace (where $\mathbf{P}_1$ is the identity). If our system has $N$ layers, we would obtain the surface GF for a system with $N$ layers,
\[
\widetilde{\mathbf{G}}_N^\mathrm{R}(E) = \left[ E \mathbf{P}_1 - \mathbf{P}_1\mathbf{HP}_1 - \mathbf{P}_1\mathbf{HQ}_1 \mathbf{G}_{\mathbf{Q}_1}(E) \mathbf{Q}_1 \mathbf{HP}_1 \right]^{-1}.
\]
The block tridiagonal structure of $\mathbf{H}$ now allows significant simplifications to this expression. First, $\mathbf{P}_1\mathbf{HQ}_1=\mathbf{H}_{1,2}\mathbf{P}_2$ because the surface layer (\textit{e.g.},\ layer 1) only couples to layer 2. A similar statement holds for $\mathbf{Q}_1\mathbf{HP}_1$. Then,
\[
\widetilde{\mathbf{G}}_N^\mathrm{R}(E) = \left[ E \mathbf{P}_1 - \mathbf{H}_{1,1} - \mathbf{H}_{1,2} \mathbf{P}_2 \mathbf{G}_{\mathbf{Q}_1}(E) \mathbf{P}_2 \mathbf{H}_{2,1} \right]^{-1}.
\]
Second, layer 2 of the total system is a surface layer of the isolated $\mathbf{Q}_1$ subsystem, meaning that $\mathbf{P}_2\mathbf{G}_{\mathbf{Q}_1}(E)\mathbf{P}_2$ is a surface GF for that subsystem, which has one fewer layer. Thus,
\[
\widetilde{\mathbf{G}}_N^\mathrm{R}(E) = \left[ E \mathbf{P}_1 - \mathbf{H}_{1,1} - \mathbf{H}_{1,2} \widetilde{\mathbf{G}}_{N-1}^\mathrm{R}(E) \mathbf{H}_{2,1} \right]^{-1}.
\]
Finally, exploiting the block Toeplitz structure and, for simplicity, replacing $\mathbf{P}_1$ by the identity, we get
\begin{equation}
\widetilde{\mathbf{G}}_N^\mathrm{R}(E) = \left[ E \mathbf{I} - \mathbf{H}_\mathrm{D} - \mathbf{H}_\mathrm{S}^\dagger \widetilde{\mathbf{G}}_{N-1}^\mathrm{R}(E) \mathbf{H}_\mathrm{S} \right]^{-1}.
\label{eq:sgf:layer-relate}
\end{equation}

Equation \eqref{eq:sgf:layer-relate} relates the surface GF for a system with $N$ layers to the surface GF for a system with $N-1$ layers. After some minor rearrangement, we can also show that a MMT describes this relationship between $\widetilde{\mathbf{G}}_N^\mathrm{R}(E)$ and $\widetilde{\mathbf{G}}_{N-1}^\mathrm{R}(E)$. Starting from Eq.\ \eqref{eq:sgf:layer-relate},
\begin{subequations}
\label{eq:sgf-mmt-r}
\begin{align}
\widetilde{\mathbf{G}}_N^\mathrm{R}(E) & = \mathbf{H}_\mathrm{S}^{-1} \left[ (E\mathbf{I} - \mathbf{H}_\mathrm{D}) \mathbf{H}_\mathrm{S}^{-1} - \mathbf{H}_\mathrm{S}^\dagger \widetilde{\mathbf{G}}_{N-1}^\mathrm{R}(E) \right]^{-1} \nonumber \\
& = \left[ \begin{array}{cc} \mathbf{0} & \mathbf{H}_\mathrm{S}^{-1} \\ -\mathbf{H}_\mathrm{S}^\dagger & (E\mathbf{I} - \mathbf{H}_\mathrm{D}) \mathbf{H}_\mathrm{S}^{-1} \end{array} \right] \bullet \widetilde{\mathbf{G}}_{N-1}^\mathrm{R}(E), \label{eq:sgf-mmt-r-onestep} \\
& = \left[ \begin{array}{cc} \mathbf{0} & \mathbf{H}_\mathrm{S}^{-1} \\ -\mathbf{H}_\mathrm{S}^\dagger & (E\mathbf{I} - \mathbf{H}_\mathrm{D}) \mathbf{H}_\mathrm{S}^{-1} \end{array} \right]^{N} \bullet \widetilde{\mathbf{G}}_{0}^\mathrm{R}(E), \label{eq:sgf-mmt-r-recur}
\end{align}
\end{subequations} 
where this last step utilizes associativity of MMTs and we define $\widetilde{\mathbf{G}}_0^\mathrm{R}(E)\equiv\mathbf{0}$. For notational convenience moving forward, we give this MMT the symbol
\[
\mathbf{M}_\mathrm{R} = \left[ \begin{array}{cc} \mathbf{0} & \mathbf{H}_\mathrm{S}^{-1} \\ -\mathbf{H}_\mathrm{S}^\dagger & (E\mathbf{I} - \mathbf{H}_\mathrm{D}) \mathbf{H}_\mathrm{S}^{-1} \end{array} \right].
\]
Similar to the transfer matrix, $\mathbf{M}_\mathrm{R}$ depends on $E$.

This process can be repeated for left surface GFs. The end result is similar,
\begin{subequations}
\label{eq:sgf-mmt-l}
\begin{align}
\widetilde{\mathbf{G}}_N^\mathrm{L}(E) & = \mathbf{M}_\mathrm{L} \bullet \widetilde{\mathbf{G}}_{N-1}^\mathrm{L}(E), \label{eq:sgf-mmt-l-onestep} \\
& = \mathbf{M}_\mathrm{L}^N \bullet \widetilde{\mathbf{G}}_{0}^\mathrm{L}(E), \label{eq:sgf-mmt-l-recur}
\end{align}
\end{subequations}
where $\widetilde{\mathbf{G}}_0^\mathrm{L}(E)\equiv\mathbf{0}$ and
\[
\mathbf{M}_\mathrm{L} = \left[ \begin{array}{cc} \mathbf{0} & \mathbf{H}_\mathrm{S}^{-\dagger} \\ -\mathbf{H}_\mathrm{S} & (E\mathbf{I} - \mathbf{H}_\mathrm{D}) \mathbf{H}_\mathrm{S}^{-\dagger} \end{array} \right].
\]

\subsection{Derivation of CBS}
\label{sec:cbs-gf:derive-cbs-gf}
From Eq.\ \eqref{eq:gf}, calculating the total system GF is tantamount to inverting the Hamiltonian. Mathematically, we thus want to consider the inverse of a block tridiagonal and block Toeplitz matrix. Theories for inverting a block tridiagonal matrix have been discussed numerous times in various contexts \cite{rozsa-447-1989, godfrin-7843-1991, meurant-707-1992, huang-7919-1997, techrpt:jain-2007, koulaei-223-2007, bk:vandebril-2008-v1}, and very recently \cite{reuter-014009-2012, boffi-015001-2015} the block Toeplitz structure with them. In the end, the block quasi-separable structure of the block tridiagonal and block Toeplitz Hamiltonian leads to a structured GF, which we now formulate. As we will soon see, surface GFs play a key role, and, from the previous discussion [Eqs.\ \eqref{eq:sgf-mmt-r} and \eqref{eq:sgf-mmt-l}], MMTs can be used to obtain all of the surface GFs. In this sense, the MMTs $\mathbf{M}_\mathrm{L}$ and $\mathbf{M}_\mathrm{R}$ are generators for $\mathbf{G}(E)$; that is, they embody the material's CBS.

To see this, let us first consider the diagonal blocks of the GF, $\mathbf{G}_{n,n}(E)$. A surface GF is one such block, but there many others unless the system is trivially small. As with the surface GFs above, we use L\"owdin partitioning \cite{lowdin-12-1963} to obtain an expression for $\mathbf{G}_{n,n}(E)$. Suppose $\mathbf{P}_n$ is the projector for our layer of interest (which is not a surface layer). Similar to before, we define $\mathbf{Q}_n=\mathbf{I}-\mathbf{P}_n$ to be the projector for all other layers of the system. A comparable statement to Eq.\ \eqref{eq:sgf:lowdin} can then be made where all $1$ subscripts are replaced by $n$. Following the same logic, we obtain an effective Hamiltonian for layer $n$ and ultimately arrive at
\begin{align*}
\mathbf{G}_{n,n}(E) & = \left[ E \mathbf{P}_n - \mathbf{P}_n\mathbf{HP}_n - \mathbf{P}_n\mathbf{HQ}_n \mathbf{G}_{\mathbf{Q}_n}(E) \mathbf{Q}_n\mathbf{HP}_n \right]^{-1} \\
& = \left[ E \mathbf{I} - \mathbf{H}_\mathrm{D} - \mathbf{P}_n\mathbf{HQ}_n \mathbf{G}_{\mathbf{Q}_n}(E) \mathbf{Q}_n\mathbf{HP}_n \right]^{-1},
\end{align*}
where we have again replaced $\mathbf{P}_n$ with the identity in the energy term.

The key difference here is that there are two disconnected parts of the $\mathbf{Q}_n$ subsystem. Layers $1,2,\ldots,n-1$ are one set, and layers $n+1,n+2,\ldots$\ are the other. Because layer $n$ only couples to layers $n-1$ and $n+1$ in the total system,
\[
\mathbf{P}_n \mathbf{HQ}_n = \mathbf{H}_{n,n-1} \mathbf{P}_{n-1} + \mathbf{H}_{n,n+1} \mathbf{P}_{n+1},
\]
and similarly for $\mathbf{Q}_n \mathbf{HP}_n$. Thus,
\begin{align*}
\mathbf{G}_{n,n}(E) = \left[ E \mathbf{I} \right. & - \mathbf{H}_\mathrm{D} \\*
& \left. - \left( \mathbf{H}_{n,n-1} \mathbf{P}_{n-1} + \mathbf{H}_{n,n+1} \mathbf{P}_{n+1} \right) \mathbf{G}_{\mathbf{Q}_n}(E) \left( \mathbf{P}_{n-1} \mathbf{H}_{n-1,n} + \mathbf{P}_{n+1}\mathbf{H}_{n+1,n} \right) \right]^{-1}.
\end{align*}
After distributing, each of these ``$\mathbf{PGP}$'' terms simplifies. First (and second), $\mathbf{P}_{n\pm1} \mathbf{G}_{\mathbf{Q}_n}(E) \mathbf{P}_{n\mp1}=\mathbf{0}$ because layers $n-1$ and $n+1$ do not belong to the same part of the $\mathbf{Q}_n$ subsystem (\textit{i.e.},\ they are decoupled in the subsystem). Third, layer $n-1$ is a surface layer for the subsystem containing layers $1,2,\ldots,n-1$; therefore (as in the surface GF discussion), $\mathbf{P}_{n-1}\mathbf{G}_{\mathbf{Q}_n}(E)\mathbf{P}_{n-1} = \widetilde{\mathbf{G}}_{n-1}^\mathrm{L}(E)$. Fourth, and similarly, layer $n+1$ is a surface layer for the subsystem containing layers $n+1,n+2,\ldots$\ such that $\mathbf{P}_{n+1}\mathbf{G}_{\mathbf{Q}_n}(E)\mathbf{P}_{n+1}=\widetilde{\mathbf{G}}_{N-n}^\mathrm{R}(E)$, where we assume there are $N$ layers in the total system. Putting these results together,
\begin{equation}
\mathbf{G}_{n,n}(E) = \left[ E \mathbf{I} - \mathbf{H}_\mathrm{D} - \mathbf{H}_\mathrm{S} \widetilde{\mathbf{G}}_{n-1}^\mathrm{L}(E) \mathbf{H}_\mathrm{S}^\dagger - \mathbf{H}_\mathrm{S}^\dagger \widetilde{\mathbf{G}}_{N-n}^\mathrm{R}(E) \mathbf{H}_\mathrm{S} \right]^{-1}.
\label{eq:gf-blocks:diagonal}
\end{equation}
Using our convention that $\widetilde{\mathbf{G}}_0^\mathrm{L/R}(E)=\mathbf{0}$, this equation also handles the cases when $n$ labels a surface layer; compare to Eq.\ \eqref{eq:sgf:layer-relate}.

It is clear from Eq.\ \eqref{eq:gf-blocks:diagonal} that calculating a diagonal block of the GF essentially reduces to calculating surface GFs. Note also that the appropriate surface GF(s) should be replaced by $\widetilde{\mathbf{G}}_\infty^\mathrm{L/R}(E)$ if the material has an infinite or semi-infinite number of layers.

\begin{figure}
\resizebox{3.25in}{!}{\includegraphics{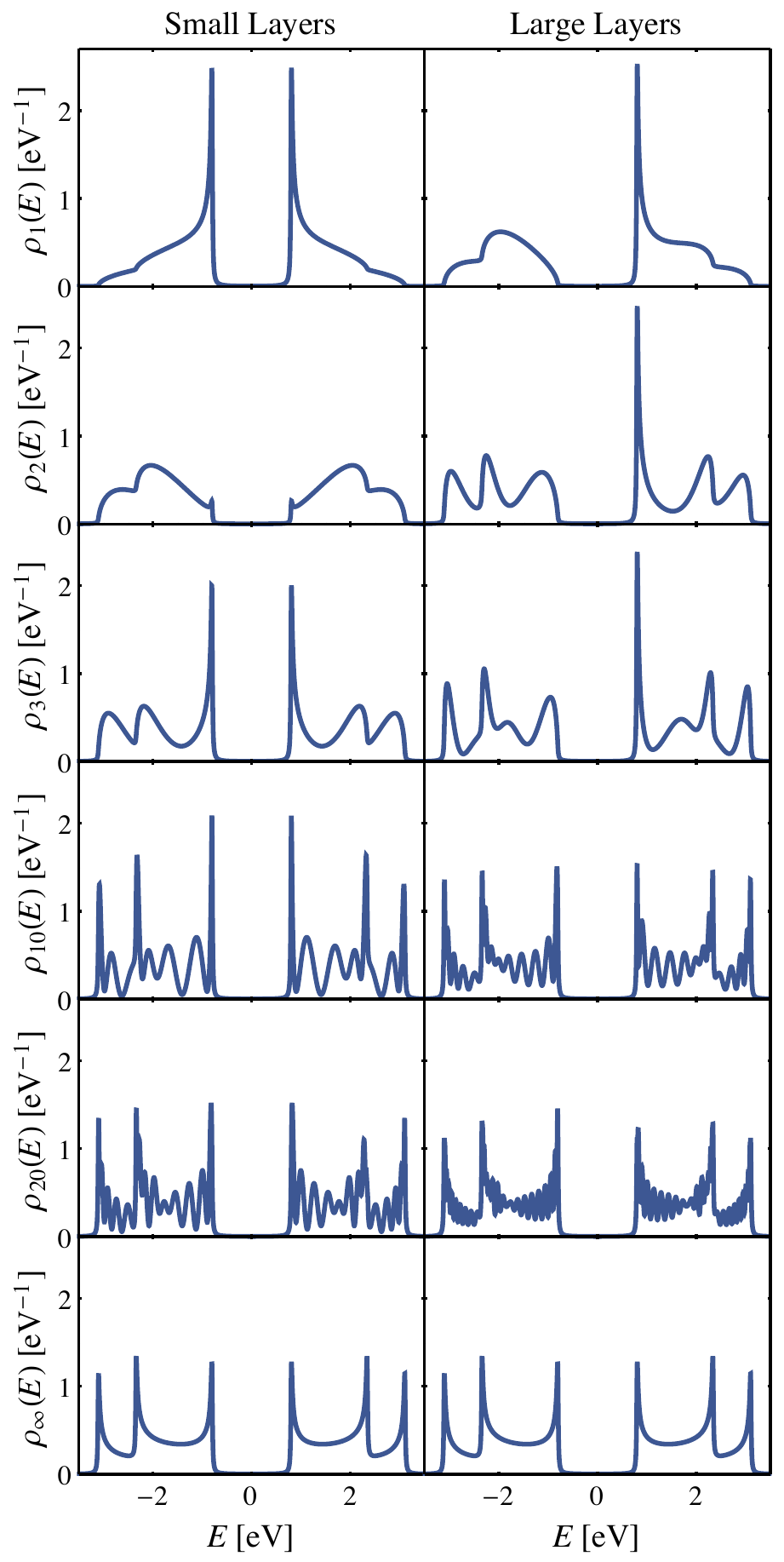}}
\caption{\label{fig:pdos}Projected DOSs, given by Eq.\ \eqref{eq:pdos}, for various layers of the example systems discussed in section \ref{sec:example}. Left and right columns: Layers chosen using the projectors from Figures \ref{fig:symmetric} and \ref{fig:asymmetric}, respectively. Each row corresponds to the projected DOS of a different layer in a semi-infinite system. From top to bottom: Layers 1, 2, 3, 10, 20, and the bulk limit (essentially layer $\infty$). Layer 1 is the surface layer and, as expected, its projected DOS is identical to the surface DOS for a semi-infinite system (see Figure \ref{fig:sgf}). Oscillations appear in the projected DOSs as we proceed from the surface to the bulk; they eventually dampen away, producing the bulk DOS, which does not depend on the choice of layer. An imaginary energy component of 10~meV was used [see Eq.\ \eqref{eq:gf}]. For comparison purposes, the projected DOSs for the large layers (right column) were scaled by a factor of $1/2$, which accounts for the different layer sizes.}
\end{figure}

Tying this result back to our running example, Figure \ref{fig:pdos} shows projected DOSs for various layers of our model system, using both choices of layers from Figures \ref{fig:symmetric} and \ref{fig:asymmetric}. Each layer's projected DOS is an instance of Eq.\ \eqref{eq:dos},
\begin{equation}
\rho_n(E) = \frac{-1}{\pi} \mathrm{Im}\left( \mathrm{Tr}\left[ \mathbf{G}_{n,n}(E) \right] \right),
\label{eq:pdos}
\end{equation}
and has been used \cite{kalkstein-85-1971, brasher-4868-1980, inglesfield-57-1987, hod-233401-2007, reuter-034703-2010, reuter-085412-2011, reuter-084707-2013, dang-155307-2014} to study quantum size effects. The projected DOS for layer 1 (\textit{i.e.},\ the surface layer) is, as expected, the surface DOS for the semi-infinite system. As we move away from the surface, oscillations appear in the projected DOSs, which eventually converge to the bulk DOS (essentially layer $\infty$) at a rate determined by CBS [point \ref{interpretations:surfaces:bulkboundary} below]. The van Hove singularities \cite{van-hove-1189-1953} at the band edges of the bulk DOS signify of a one-dimensional material. Although the choice of layer is critically important for surface-related properties, it does not matter for bulk properties.

Let us now consider the off-diagonal blocks of the GF. Expressions for these blocks can be straightforwardly obtained from Dyson's equation \cite{bk:economou-2006}, which, similar to L\"owdin partitioning, has us break up the system. Unlike L\"owdin partitioning, one of our ``partitions'' will only contain the coupling between two layers; the other ``partition'' contains everything else. The idea is best presented through example. Suppose we want to calculate $\mathbf{G}_{n+1,n}(E)$. We regard the coupling between layers $n$ and $n+1$,
\[
\mathbf{V} = \mathbf{P}_n\mathbf{HP}_{n+1} + \mathbf{P}_{n+1}\mathbf{HP}_n,
\]
as a perturbation on the Hamiltonian $\mathbf{H}_0$, which describes layers $1,2,\ldots,n$ decoupled from layers $n+1,n+2,\ldots$. In other words, $\mathbf{H}_0=\mathbf{H}-\mathbf{V}$. Dyson's equation relates the GF for the total system ($\mathbf{G}$) to the GF for the unperturbed system ($\mathbf{G}_0$),
\begin{equation}
\mathbf{G}(E) = \mathbf{G}_0(E) + \mathbf{G}(E)\mathbf{V}\mathbf{G}_0(E).
\label{eq:dyson}
\end{equation}
Applying this to the $(n+1,n)$ block,
\[
\mathbf{G}_{n+1,n}(E) = \mathbf{P}_{n+1}\mathbf{G}_0(E)\mathbf{P}_n + \mathbf{P}_{n+1}\mathbf{G}(E) \left( \mathbf{P}_n\mathbf{HP}_{n+1} + \mathbf{P}_{n+1}\mathbf{HP}_n \right) \mathbf{G}_0(E)\mathbf{P}_n.
\]
Because layers $n$ and $n+1$ are decoupled in $\mathbf{H}_0$, $\mathbf{P}_{n+1}\mathbf{G}_0(E)\mathbf{P}_n=\mathbf{0}$. Layer $n$ is also a surface layer in $\mathbf{H}_0$ such that $\mathbf{P}_n\mathbf{G}_0(E)\mathbf{P}_n=\widetilde{\mathbf{G}}_n^\mathrm{L}(E)$. Thus,
\begin{align*}
\mathbf{G}_{n+1,n}(E) & = \mathbf{P}_{n+1}\mathbf{G}(E) \mathbf{P}_{n+1}\mathbf{HP}_n \mathbf{G}_0(E)\mathbf{P}_n \\
& = \mathbf{G}_{n+1,n+1}(E) \mathbf{H}_\mathrm{S} \widetilde{\mathbf{G}}_n^\mathrm{L}(E).
\end{align*}

This procedure of isolating the coupling between two layers and applying Dyson's equation can be iterated, leading to the general result
\begin{subequations}
\label{eq:gf-blocks}
\begin{equation}
\mathbf{G}_{m,n}(E) = \mathbf{G}_{m,n+1}(E)\mathbf{H}_\mathrm{S}\widetilde{\mathbf{G}}_n^\mathrm{L}(E) \text{ when } m>n.
\label{eq:gf-blocks:subleft}
\end{equation}
Similar logic can be used when $m<n$, and results in
\begin{equation}
\mathbf{G}_{m,n}(E) = \mathbf{G}_{m,n-1}(E) \mathbf{H}_\mathrm{S}^\dagger \widetilde{\mathbf{G}}_{N-n+1}^\mathrm{R}(E) \text{ when } m < n.
\label{eq:gf-blocks:superright}
\end{equation}
Finally, Dyson's equation can equivalently be written as $\mathbf{G}(E) = \mathbf{G}_0(E) + \mathbf{G}_0(E)\mathbf{V}\mathbf{G}(E)$, which produces the equivalent expressions
\begin{align}
\mathbf{G}_{m,n}(E) & = \widetilde{\mathbf{G}}_{N-m+1}^\mathrm{R}(E) \mathbf{H}_\mathrm{S} \mathbf{G}_{m-1,n}(E) \text{ when } m > n, \label{eq:gf-blocks:subright} \\
\mathbf{G}_{m,n}(E) & = \widetilde{\mathbf{G}}_m^\mathrm{L}(E) \mathbf{H}_\mathrm{S}^\dagger \mathbf{G}_{m+1,n}(E) \text{ when } m < n. \label{eq:gf-blocks:superleft}
\end{align}
\end{subequations}
These equations can also be extended to infinite or semi-infinite systems by replacing the appropriate surface GFs with $\widetilde{\mathbf{G}}_\infty^\mathrm{L/R}(E)$.
 
Equations\ \eqref{eq:gf-blocks:diagonal} and \eqref{eq:gf-blocks} demonstrate that every block of the GF can be straightforwardly obtained from surface GFs. Then, because the surface GFs are described by the MMTs $\mathbf{M}_\mathrm{L}$ and $\mathbf{M}_\mathrm{R}$, these MMTs transitively generate the entire GF \cite{boffi-015001-2015}.

This last statement suggests that $\mathbf{M}_\mathrm{L}$ and $\mathbf{M}_\mathrm{R}$ access the material's CBS in a way that facilitates matrix inversion; that is, $\mathbf{M}_\mathrm{L}$ and $\mathbf{M}_\mathrm{R}$ provide a GF-based route to CBS. We demonstrate this by showing some common structure between these MMTs and the transfer matrix $\mathbf{T}$ in Eq.\ \eqref{eq:transfer-matrix}. Mathematically, it is straightforward to prove that $\mathbf{M}_\mathrm{L}$ is similar to $\mathbf{T}$,
\[
\mathbf{M}_\mathrm{L} = \left[ \begin{array}{cc} \mathbf{0} & \mathbf{I} \\ \mathbf{H}_\mathrm{S}^\dagger & \mathbf{0} \end{array} \right] \mathbf{T} \left[ \begin{array}{cc} \mathbf{0} & \mathbf{H}_\mathrm{S}^{-\dagger} \\ \mathbf{I} & \mathbf{0} \end{array} \right];
\]
meaning that $\mathbf{M}_\mathrm{L}$ has the same eigenvalues as $\mathbf{T}$. From section \ref{sec:cbs-wf}, these eigenvalues are essentially the material's CBS. Likewise, $\mathbf{M}_\mathrm{R}$ is similar to $\mathbf{T}^{-1}$,
\[
\mathbf{M}_\mathrm{R} = \left[ \begin{array}{cc} \mathbf{I} & \mathbf{0} \\ \mathbf{0} & \mathbf{H}_\mathrm{S} \end{array} \right] \mathbf{T}^{-1} \left[ \begin{array}{cc} \mathbf{I} & \mathbf{0} \\ \mathbf{0} & \mathbf{H}_\mathrm{S}^{-1} \end{array} \right].
\]
In other words, $\lambda=e^{ika}$ is an eigenvalue of $\mathbf{T}$ (or $\mathbf{M}_\mathrm{L}$) if and only if $1/\lambda=e^{-ika}$ is an eigenvalue of $\mathbf{M}_\mathrm{R}$. For reference,
\[
\mathbf{T}^{-1} = \left[ \begin{array}{cc} \mathbf{0} & \mathbf{I} \\ - \mathbf{H}_\mathrm{S}^{-1} \mathbf{H}_\mathrm{S}^\dagger & \mathbf{H}_\mathrm{S}^{-1}(E \mathbf{I} - \mathbf{H}_\mathrm{D}) \end{array} \right].
\]
This relationship between $\mathbf{T}$, $\mathbf{M}_\mathrm{L}$, and $\mathbf{M}_\mathrm{R}$ is the multi-band generalization of a likewise statement for one-band materials, where $\mathbf{T}$, $\mathbf{M}_\mathrm{L}$, and $\mathbf{M}_\mathrm{R}$ are $2\times 2$ matrices and complex M\"obius transformations were used \cite{sanchez-soto-191-2012}.

\section{Mathematical Properties of Complex Band Structure}
\label{sec:properties}

Equipped with insight from the example in section \ref{sec:example} and these wavefunction- and GF-based routes to CBS, we now discuss some mathematical properties of CBS. Many of the results in this section and the next have been reported in the various derivations of CBS. For brevity, we do not necessarily cite and/or compare every development of the results, but instead focus on the results themselves. Each point will first state the main result. In most cases, more information, and possibly justification, will then be given.

\subsection{The Dispersion Relation}
\label{sec:properties:general}
There are a handful of properties that apply regardless of the material's dimensionality. More detailed discussions of these results can be found in Refs.\ \onlinecite{blount-305-1962, heine-300-1963, krieger-776-1967, prodan-035128-2006}.

\begin{enumerate}[label=(\ref{sec:properties}.\arabic*),align=left,series=properties]
\item \label{properties:general:analytic-dispersion} The dispersion relation $E(\vec{k})$ is analytic \cite{bk:needham-1997} except at $\vec{k}$ where multiple bands intersect (\textit{i.e.},\ branch points). Proofs are presented in Refs.\ \onlinecite{blount-305-1962, des-cloizeaux-a685-1964, krieger-776-1967, prodan-035128-2006}.

\item \label{properties:general:reciprocal} If $\vec{g}$ is a reciprocal lattice vector (and thus real), then $E(\vec{k})=E(\vec{k} + \vec{g})$. This result is easily verified by using Bloch's theorem in the time-independent Schr\"odinger equation.

\item \label{properties:general:conjugation} $E(\vec{k})=E(\vec{k}^\ast)^\ast$, where $^\ast$ denotes complex conjugation. Similar to the previous point, this result can also be verified using Bloch's theorem in the time-independent Schr\"odinger equation. Furthermore, if $E$ is real, then both $\vec{k}$ and $\vec{k}^\ast$ are in the material's CBS. This symmetry between $\vec{k}$ and $\vec{k}^\ast$ can be observed in Figures \ref{fig:symmetric} and \ref{fig:asymmetric} from our example.

\item \label{properties:general:bp} Branch points occur at non-real $\vec{k}$ when the potential (and thus Hamiltonian) are Hermitian \cite{prodan-035128-2006}. However, branch points can be shifted to real $\vec{k}$ if the potential is non-Hermitian, as might happen in (for example) high-energy electron diffraction \cite{buxton-3941-1977}.
\end{enumerate}

\subsection{Heine's ``Lines of Real Energy''}
\label{sec:properties:realE}
Heine \cite{heine-300-1963} derived several results about the ``lines of real $E$'' when $\vec{k}$ is only complex in one dimension that were subsequently classified by Chang \cite{chang-605-1982}. These results are all visible in our example from section \ref{sec:example} (Figures \ref{fig:symmetric} and \ref{fig:asymmetric}), which may be a useful companion.

\begin{enumerate}[resume*=properties]
\item \label{properties:realE:winding} Away from branch points and band edges, lines of real energy do not merge, split, begin, or terminate. Heine \cite{heine-300-1963} demonstrated these results by considering the winding number \cite{bk:needham-1997} of the function $E(k)-E(k_0)$ for some wavevector $k_0$ that is not a branch point or band edge. There is only one zero of this function inside some infinitesimally-small, closed contour around $k_0$, meaning that the function accumulates a phase of $2\pi$ going around the contour. Consequently, there are two points with real energies on the contour: One for the line to enter and one for it to leave. If a line merged, split, began, or terminated, there would have to be an odd number of real energies on the contour.

\item \label{properties:realE:bandedge} At a band edge, two lines of real energy meet and take $90^\circ$ turns into the complex plane. A proof is straightforward. Suppose $k_0$ is a band edge. $E(k)$ has an extremum along the real axis at $k_0$, meaning $E'(k_0)=0$. Then, because $E(k)$ is analytic at $k_0$, we know that
\begin{align*}
E(k) & = E(k_0) + E'(k_0) (k-k_0) + \frac{E''(k_0)}{2} (k-k_0)^2 + \mathcal{O}[(k-k_0)^3] \\
& \approx E(k_0) + \frac{E''(k_0)}{2} (k-k_0)^2
\end{align*}
for $k$ near $k_0$. Because $k_0$ is in a real band, $E(k_0)$ and $E''(k_0)$ are both real, implying that $(k-k_0)^2$ must be real for $E(k)$ to be real. Thus, $k-k_0$ is either real (meaning $k$ is also in the conventional band) or purely imaginary (corresponding to a $90^\circ$ turn). From this reasoning, we also see that a band edge $k_0$ is a saddle point of $E(k)$.

\item \label{properties:realE:branchpoint} Lines of real energy do not generally go through branch points. In other words, if $k_\mathrm{bp}$ is a branch point, $E(k_\mathrm{bp})$ is not usually real. See the example in Figure \ref{fig:asymmetric}. The one major exception \cite{heine-300-1963} to this statement occurs when there is a mirror plane between between the layers, as demonstrated in Figure \ref{fig:symmetric}. $E(k_\mathrm{bp})$ will be real in this case. Either way \cite{heine-300-1963},
\begin{equation}
E(k) \approx E(k_\mathrm{bp}) + C \left( k - k_\mathrm{bp} \right)^{1/2},
\label{eq:dispersion-bp}
\end{equation}
for $k$ near $k_\mathrm{bp}$ and some (generally complex) constant $C$. Note that Eq.\ \eqref{eq:dispersion-bp} is only valid in the likely case that two bands meet at the branch point [see point \ref{interpretations:general:bp}]. If $m$ bands were to intersect at $k_\mathrm{bp}$, the exponent $1/2$ would be replaced by $1/m$ \cite{krieger-776-1967}.
\end{enumerate}

\subsection{One-Dimensional Complex Band Structure}
\label{sec:properties:1d}
Because one-dimensional CBS is related to the eigenvalues of the transfer matrix $\mathbf{T}$ (alternatively and equivalently, the MMTs $\mathbf{M}_\mathrm{L}$ and $\mathbf{M}_\mathrm{R}$), we focus on the spectral decomposition of $\mathbf{T}$.

\begin{enumerate}[resume*=properties]
\item \label{properties:1d:nonhermitian} The transfer matrix is not Hermitian. As desired, this means that $\mathbf{T}$ can have complex eigenvalues.

\item \label{properties:1d:pairedevals} The eigenvalues of $\mathbf{T}$ come in pairs: If $\lambda$ is an eigenvalue, so is $1/\lambda^\ast$. To see this, we verify that $\mathbf{T}$ is similar to $\mathbf{T}^{-\dagger}$ (mathematically, $\mathbf{T}$ is complex symplectic),
\[
\mathbf{T} = \left[ \begin{array}{cc} \mathbf{0} & -\mathbf{H}_\mathrm{S}^{-\dagger} \\ \mathbf{H}_\mathrm{S}^{-1} & \mathbf{0} \end{array} \right] \mathbf{T}^{-\dagger} \left[ \begin{array}{cc} \mathbf{0} & \mathbf{H}_\mathrm{S} \\ -\mathbf{H}_\mathrm{S}^\dagger & \mathbf{0} \end{array} \right].
\]

\item \label{properties:1d:evaldistribution} $\mathbf{T}$ has an even (possibly zero) number of eigenvalues with unit magnitude, and an equal number of eigenvalues with magnitudes greater than and less than 1. As follows from Fact 2.14.13 of Ref.\ \onlinecite{bk:bernstein-2009},
\[
\mathrm{det}\left(\mathbf{T}\right)=\mathrm{det}\left(\mathbf{H}_\mathrm{S}^{-\dagger}\mathbf{H}_\mathrm{S}\right) = \frac{\mathrm{det}(\mathbf{H}_\mathrm{S})}{\mathrm{det}(\mathbf{H}_\mathrm{S})^\ast}.
\]
Thus, $|\mathrm{det}(\mathbf{T})|=1$, meaning the product of all eigenvalues is 1 (in magnitude). Appealing to the eigenvalue pairing in point \ref{properties:1d:pairedevals}, it is straightforward to see the conclusion. Suppose that $\lambda$ is an eigenvalue of $\mathbf{T}$ with unit magnitude. Then, $1/\lambda^\ast$ has unit magnitude and is also an eigenvalue. Similar reasoning reveals that there are an equal number of eigenvalues with $|\lambda|<1$ and $|\lambda|>1$.

\item \label{properties:1d:eval1realE} Eigenvalues with unit magnitude [see points \ref{interpretations:1d:propagate} and \ref{interpretations:1d:evanescent} below] are forbidden when $E$ is not real. This situation arises, for example, when numerically evaluating the limit in Eq.\ \eqref{eq:gf} and a small imaginary component is present in the energy. In essence, the imaginary energy breaks symmetry so that the paired eigenvalues are no longer degenerate. Half of the states will have $|\lambda|<1$ and the other half will have $|\lambda|>1$ \cite{velev-r637-2004}. Note that, as expected, $|\lambda|\to1$ as $\mathrm{Im}(E)\to0^\pm$ for these eigenvalues.

A proof of this result is straightforward and proceeds by contradiction. Suppose that $\lambda$ is an eigenvalue of $\mathbf{T}$ ($|\lambda|=1$) when $\mathrm{Im}(E)\neq0$. Then, from Eq.\ \eqref{eq:transfer-matrix} and Fact 2.14.13 in Ref.\ \onlinecite{bk:bernstein-2009},
\begin{eqnarray}
0 & = & \mathrm{det} \left[ \lambda \mathbf{I} - \mathbf{T} \right] \nonumber \\
& = & \mathrm{det} \left[ \begin{array}{cc} \lambda \mathbf{I} - \mathbf{H}_\mathrm{S}^{-\dagger} (E\mathbf{I} - \mathbf{H}_\mathrm{D}) & \mathbf{H}_\mathrm{S}^{-\dagger} \mathbf{H}_\mathrm{S} \\ -\mathbf{I} & \lambda \mathbf{I} \end{array} \right] \nonumber \\
& = & \mathrm{det} \left[ -\lambda \mathbf{H}_\mathrm{S}^{-\dagger} \right] \mathrm{det} \left[ E \mathbf{I} - \lambda \mathbf{H}_\mathrm{S}^\dagger - \mathbf{H}_\mathrm{D} - \frac{1}{\lambda} \mathbf{H}_\mathrm{S} \right]. \nonumber
\end{eqnarray}
The first determinant is nonzero because $|\lambda|=1$ and $\mathbf{H}_\mathrm{S}^{-\dagger}$ is invertible. Considering the second determinant, we can write $\lambda=e^{i\theta}$ ($\theta$ is real) such that
\begin{equation}
0 = \mathrm{det} \left[ E \mathbf{I} - e^{i\theta} \mathbf{H}_\mathrm{S}^\dagger - \mathbf{H}_\mathrm{D} - e^{-i\theta} \mathbf{H}_\mathrm{S} \right].
\label{eq:proof-lambda}
\end{equation}
Recalling that $\mathbf{H}_\mathrm{D}$ is Hermitian, it is easily verified that $e^{i\theta} \mathbf{H}_\mathrm{S}^\dagger + \mathbf{H}_\mathrm{D} + e^{-i\theta} \mathbf{H}_\mathrm{S}$ is also Hermitian. Then, because Eq.\ \eqref{eq:proof-lambda} can only hold when $E$ is an eigenvalue of $e^{i\theta} \mathbf{H}_\mathrm{S}^\dagger + \mathbf{H}_\mathrm{D} + e^{-i\theta} \mathbf{H}_\mathrm{S}$, we reach a contradiction. Hermitian operators must have real eigenvalues and $E$ is not real. Thus, eigenvalues of $\mathbf{T}$ cannot have unit magnitude when $E$ is not real.

\item \label{properties:1d:diagonalizable} $\mathbf{T}$ may not be diagonalizable. In general, the transfer matrix is not normal, meaning $\mathbf{TT}^\dagger\neq\mathbf{T}^\dagger\mathbf{T}$. Consequently, $\mathbf{T}$ is not guaranteed to have a complete set of eigenvectors and any eigenvectors it does possess are not required to be mutually orthogonal \cite{bk:golub-2012}. Rather than diagonalizing $\mathbf{T}$, a Jordan or Schur decomposition \cite{bk:golub-2012} (which both generalize the notion of ``diagonalization'') may be more appropriate for analytical or numerical considerations, respectively. Additional structure of the eigenvectors can be found in Ref.\ \onlinecite{biczo-1992-1985}.
\end{enumerate}

\section{Physical Interpretations and Applications of Complex Band Structure}
\label{sec:interpretations}

The mathematical properties of CBS discussed in the last section set the stage for physical interpretations of CBS, which we now discuss. As before, we first state the main result for each point and then offer some discussion or justification. Some of the justifications are detailed (particularly those in Sec.\ \ref{sec:interpretations:surfaces}), but are not necessary for proceeding to subsequent points. We provide these details as references for the interested reader.

\subsection{The Dispersion Relation}
\label{sec:interpretations:general}
\begin{enumerate}[label=(\ref{sec:interpretations}.\arabic*),align=left,series=interpretations]
\item \label{interpretations:general:branch} Each branch of the multi-valued dispersion relation corresponds to a band of the material. In this context, a ``band'' includes the real band from conventional band structure, as well as some complex wavevectors. Prodan \cite{prodan-035128-2006} further discusses, with examples of the various Riemann surfaces, branches and branch points of $E(k)$ when $k$ is strictly one-dimensional.

\item \label{interpretations:general:bp} Branch points of $E(\vec{k})$ occur at the boundaries between two bands. Consequently, they will only appear in band gaps. It may be possible that branch points will appear inside a band for higher-dimensional (not one-dimensional) materials \cite{mahboob-201307r-2004, schleife-012014-2009}; however, in the absence of general tools for higher-dimensional CBS, heuristics were used to calculate those branch point energies \footnote{Heuristic branch point energies for higher-dimensional materials were obtained by averaging over a heuristic branch point energy for each real wavevector in the material's first Brillouin zone. It is possible that this average is above the lower band edge (which comes from a specific $\vec{k}$) of the conduction band or below the upper band edge of the valence band.}.

\item \label{interpretations:general:character} If $\vec{k}_\mathrm{bp}$ is a branch point [and therefore in a band gap from point \ref{interpretations:general:bp}], states near $\vec{k}_\mathrm{bp}$ with $E<E(\vec{k}_\mathrm{bp})$ will have more ``valence band character'', and similarly for ``conduction band character'' when $E>E(\vec{k}_\mathrm{bp})$ \cite{appelbaum-4973-1974, rehr-448-1974, rehr-1981-1974}. Using chemistry nomenclature, states near the branch point with $E=E(\vec{k}_\mathrm{bp})$ are ``non-bonding''.
\end{enumerate}

\subsection{One-Dimensional Complex Band Structure}
\label{sec:interpretations:1d}

\begin{enumerate}[resume*=interpretations]
\item \label{interpretations:1d:propagate} Eigenvalues of $\mathbf{T}$ with unit magnitude correspond to the bulk-propagating states from conventional band structure. Briefly, if $\lambda=e^{ika}$ and $|\lambda|=1$, then $\mathrm{Im}(k)=0$. Because $k$ is real, $\lambda$ and $1/\lambda^\ast$ are degenerate eigenvalues. If $E$ is outside of a band, there will not be any of these states or eigenvalues.

\item \label{interpretations:1d:evanescent} Eigenvalues of $\mathbf{T}$ with non-unit magnitudes correspond to the evanescent states that are excluded from conventional band structure. Depending on orientation, states corresponding to $|\lambda|<1$ grow or decay from one layer to the next, whereas states with $|\lambda|>1$ exhibit the opposite behavior. From point \ref{properties:1d:evaldistribution}, there are an equal number of states with $|\lambda|>1$ and $|\lambda|<1$.

\item \label{interpretations:1d:defective} The states corresponding to defective eigenvalues of $\mathbf{T}$ (\textit{i.e.},\ the degenerate eigenvalues that lack a complete set of eigenvectors, thereby prohibiting $\mathbf{T}$ from being diagonalized) are not well understood. Some results \cite{blount-305-1962, cottey-1235-1971, cottey-2583-1972, cottey-2591-1972, buxton-3941-1977, biczo-31-1979, biczo-347-1981, biczo-1992-1985, biczo-503-1998, reuter-034703-2010, reuter-084707-2013} suggest that $\mathbf{T}$ is defective at a band edge or a branch point, or when a surface state exists [point \ref{interpretations:surfaces:states}]. More work needs to be performed to understand the significance of defective $\mathbf{T}$.

\item \label{interpretations:1d:dynamics} CBS determines the dynamic properties of electrons in a material \cite{des-cloizeaux-a685-1964}, which are described by the off-diagonal blocks of the GF. To see this, we apply general properties of MMTs \cite{boffi-015001-2015}. Equation \eqref{eq:gf-blocks:subright} takes a lower-triangular block of the GF and moves one layer down, and can be approximated as
\[
\mathbf{G}_{j,k}(E) \approx \widetilde{\mathbf{G}}_\infty^\mathrm{R}(E) \mathbf{H}_\mathrm{S} \mathbf{G}_{j-1,k}(E),
\]
when $j$ is sufficiently large [see point \ref{interpretations:surfaces:converge} below]. Then, using Eq.\ \eqref{eq:gf-fp},
\begin{eqnarray}
\mathbf{G}_{j,k}(E) & = & \left( \mathbf{M}_\mathrm{R} \bullet \widetilde{\mathbf{G}}_\infty^\mathrm{R}(E) \right) \mathbf{H}_\mathrm{S} \mathbf{G}_{j-1,k}(E) \nonumber \\
& = & \mathbf{H}_\mathrm{S}^{-1} \left[ -\mathbf{H}_\mathrm{S}^\dagger \widetilde{\mathbf{G}}_\infty^\mathrm{R}(E) + (E\mathbf{I} - \mathbf{H}_\mathrm{D}) \mathbf{H}_\mathrm{S}^{-1} \right]^{-1} \mathbf{H}_\mathrm{S} \mathbf{G}_{j-1,k}(E). \nonumber
\end{eqnarray}
We now invoke a general property of MMTs (Thm.\ 2 of Ref.\ \onlinecite{boffi-015001-2015}): The eigenvalues of $-\mathbf{H}_\mathrm{S}^\dagger \widetilde{\mathbf{G}}_\infty^\mathrm{R}(E) + (E\mathbf{I} - \mathbf{H}_\mathrm{D}) \mathbf{H}_\mathrm{S}^{-1}$ are exactly the eigenvalues of $\mathbf{M}_\mathrm{R}$ with $|\lambda|>1$ [assuming $\mathrm{Im}(E)\neq0$]. In this sense, the material's CBS determines the behavior of the GF as we move off the diagonal; these are the dynamical properties of electrons. If $E$ is outside of a band, the electron transition probabilities fall off rapidly with each additional layer because there are no propagating states. Conversely, if $E$ is in a band, there is at least one eigenvalue satisfying $|\lambda|\approx1$, and the off-diagonal blocks of the GF are slow to decay. As a final comment, a similar analysis can be performed for each part of Eq.\ \eqref{eq:gf-blocks}, and leads to the same result.
\end{enumerate}

\subsection{Surfaces}
\label{sec:interpretations:surfaces}
Surfaces were one of the first applications of CBS. Notably, and as discussed in section \ref{sec:cbs-gf}, CBS is closely linked to the surface GF for a semi-infinite system \cite{garcia-moliner-1789-1969, dy-4237-1979, lee-4997-1981, lee-355-1981, umerski-5266-1997, velev-r637-2004, rungger-035407-2008, reuter-085412-2011, bravi-155445-2014}. Herein we describe some additional results related to surfaces.

\begin{enumerate}[resume*=interpretations]
\item \label{interpretations:surfaces:fp} The surface GF (for a semi-infinite system) can be directly obtained from the eigenvectors of the transfer matrix \cite{lee-4997-1981} or the MMTs $\mathbf{M}_\mathrm{L}$ and $\mathbf{M}_\mathrm{R}$ \cite{umerski-5266-1997, reuter-085412-2011}; see Eq.\ \eqref{eq:surface-gf-mmt}. (Generalized eigenvectors may also be needed if the transfer matrix is not diagonalizable.) This explains why the surface GFs in Figure \ref{fig:sgf} are different for the two choices of layers. Even though the eigenvalues of $\mathbf{M}_\mathrm{R}$, $\mathbf{M}_\mathrm{L}$, or $\mathbf{T}$ (\textit{i.e.},\ the dispersion relation) are mostly independent of the choice of layer (see Figures \ref{fig:symmetric} and \ref{fig:asymmetric}), the eigenvectors are much more sensitive.

To elaborate, consider a semi-infinite material with (right) surface GF $\widetilde{\mathbf{G}}_\infty^\mathrm{R}(E)$. When applied to a  surface GF, the MMT $\mathbf{M}_\mathrm{R}$ produces the surface GF for a system with an additional layer. This new system, however, is identical to the original semi-infinite system; thus,
\begin{equation}
\widetilde{\mathbf{G}}_\infty^\mathrm{R}(E) = \mathbf{M}_\mathrm{R} \bullet \widetilde{\mathbf{G}}_\infty^\mathrm{R}(E).
\label{eq:gf-fp}
\end{equation}
Mathematically, the surface GF for a semi-infinite system is a fixed point of $\mathbf{M}_\mathrm{R}$.

We previously showed \cite{boffi-015001-2015} that the fixed points of a MMT are related to the MMT's invariant subspaces \cite{bk:gohberg-2006}. In other words, a fixed point corresponds to a collection of eigenvectors of the MMT (and possibly generalized eigenvectors if the MMT is not diagonalizable). Physically, we are interested in the half of the eigenvectors that correspond to right- (or left-)propagating states; that is, those with eigenvalues greater than (or less than) 1 in magnitude. Note that, if there are eigenvalues with unit magnitude (\textit{i.e.},\ $E$ is in a band) such that half of the eigenvalues are not greater (less) than 1 in magnitude [point \ref{properties:1d:pairedevals}], it is well known that the surface GF is not well-defined. Adding a small imaginary part to the energy [point \ref{properties:1d:eval1realE}] eliminates this problem.

A proof of the upcoming result is lengthy and requires other properties of MMTs that we will not discuss here. For brevity, we simply state the result and direct the interested reader to Ref.\ \onlinecite{boffi-015001-2015} for more details. We also have to impose a basis set and assume that $\mathbf{H}_\mathrm{D}$ and $\mathbf{H}_\mathrm{S}$ are accurately represented by $M\times M$ matrices \footnote{The requirement of a basis in point \ref{interpretations:surfaces:fp} can be lifted by appealing to the projective space interpretation of a MMT. More details can be found in Ref.\ \onlinecite{boffi-015001-2015}.}. Let $\mathbf{U}$ be the $2M\times M$ matrix whose columns are the $M$ (generalized) eigenvectors of $\mathbf{M}_\mathrm{R}$ with eigenvalues greater than 1 in magnitude (recall that $\mathbf{M}_\mathrm{R}$ will be a $2M\times 2M$ matrix). Then,
\begin{equation}
\widetilde{\mathbf{G}}_\infty^\mathrm{R}(E) = \mathbf{U}_1 \mathbf{U}_2^{-1},
\label{eq:surface-gf-mmt}
\end{equation}
where $\mathbf{U}_1$ is the first $M$ rows of $\mathbf{U}$ and $\mathbf{U}_2$ is the bottom $M$ rows of $\mathbf{U}$. Likewise, if $\mathbf{V}$ is the $2M\times M$ matrix of eigenvectors of $\mathbf{M}_\mathrm{L}$ with eigenvalues greater than 1 in magnitude, then
\[
\widetilde{\mathbf{G}}_\infty^\mathrm{L}(E) = \mathbf{V}_1 \mathbf{V}_2^{-1},
\]
where $\mathbf{V}_1$ and $\mathbf{V}_2$ are similarly defined.

\item \label{interpretations:surfaces:converge} CBS determines the rate at which $\widetilde{\mathbf{G}}_N^\mathrm{L/R}(E)$ converges to $\widetilde{\mathbf{G}}_\infty^\mathrm{L/R}(E)$ with increasing $N$ (if it converges), which is part of an examination of quantum size effects. This result is also relevant to numerical methods that approximate $\widetilde{\mathbf{G}}_\infty^\mathrm{L/R}(E)$ by $\widetilde{\mathbf{G}}_N^\mathrm{L/R}(E)$ for $N$ sufficiently large \cite{lopez-sancho-851-1985, velev-r637-2004}. Specifically, $\widetilde{\mathbf{G}}_N^\mathrm{L/R}(E)$ converges to $\widetilde{\mathbf{G}}_\infty^\mathrm{L/R}(E)$ as $\mathcal{O}(|\lambda_M|^{2N})$, where $\lambda_M$ is the largest (in magnitude) eigenvalue of $\mathbf{M}_\mathrm{L/R}$ satisfying $|\lambda_M|\le1$. As implied, $\widetilde{\mathbf{G}}_N^\mathrm{L/R}(E)$ does not converge to the semi-infinite limit when $|\lambda_M|=1$. In this case, $E$ is in a band and $\widetilde{\mathbf{G}}_\infty^\mathrm{L/R}(E)$ is not well-defined.

Starting from Eq.\ \eqref{eq:sgf-mmt-r-recur} and using the Jordan form of $\mathbf{M}_\mathrm{R}=\mathbf{U}_\mathrm{R}\mathbf{J}\mathbf{U}_\mathrm{R}^{-1}$,
\[
\widetilde{\mathbf{G}}_N^\mathrm{R}(E) = \left( \mathbf{U}_\mathrm{R} \mathbf{J}^N \mathbf{U}_\mathrm{R}^{-1} \right) \bullet \widetilde{\mathbf{G}}_0^\mathrm{R}(E).
\]
Similar to the discussion in point \ref{interpretations:surfaces:fp}, we assume that $\mathrm{Im}(E)>0$ such that half of the eigenvalues in $\mathbf{J}$ are greater than 1 (in magnitude) and the other half are less than 1 (in magnitude); see points \ref{properties:1d:evaldistribution} and \ref{properties:1d:eval1realE}. We can then order the eigenvalues in $\mathbf{J}$ so that each element of $\mathbf{U}_\mathrm{R}^{-1}\bullet \widetilde{\mathbf{G}}_0^\mathrm{R}(E)$ is multiplied by an eigenvalue that is smaller than 1 (in magnitude) and simultaneously divided by an eigenvalue that is larger than 1 (in magnitude) when $\mathbf{J}^N$ is applied. Consequently, each element of $(\mathbf{J}^{N}\mathbf{U}_\mathrm{R}^{-1})\bullet \widetilde{\mathbf{G}}_0^\mathrm{R}(E)$ exponentially converges to $0$ with a rate determined by CBS. The slowest element to approach zero converges as $\mathcal{O}(|\lambda_M|^{2N})$.

\item \label{interpretations:surfaces:states} CBS and the (semi-infinite) surface GF detail the existence of surface states and their asymptotic decay rates. Surface states appear at energies where there are no bulk-propagating states (that is, no $\lambda$ have unit magnitude) and $\widetilde{\mathbf{G}}_\infty^\mathrm{L/R}(E)$ has a pole \cite{dy-4237-1979, lee-4988-1981}. The latter condition is satisfied when, mathematically (with a momentary abuse of notation),
\[
0 = \mathrm{det}\left[ \left( \widetilde{\mathbf{G}}_\infty^\mathrm{L/R}(E) \right)^{-1} \right].
\]
From Eq.\ \eqref{eq:surface-gf-mmt}, this condition rigorously becomes
\begin{equation}
0 = \mathrm{det}\left( \mathbf{U}_2 \right) \text{ or } 0 = \mathrm{det}\left( \mathbf{V}_2 \right),
\label{eq:surface-state-condition}
\end{equation}
with $\mathbf{U}_2$ and $\mathbf{V}_2$ defined as in point \ref{interpretations:surfaces:fp}.
Last, if there exists a surface state at energy $E$, the state asymptotically decays (in magnitude) at a rate of $-\ln|\lambda_{M}|$ per layer \cite{lee-4997-1981}, where $\lambda_M$ is the largest eigenvalue (in magnitude) satisfying $|\lambda_M|<1$. Physically, $\lambda_M$ corresponds to the most slowly decaying evanescent state from the material's CBS.

\item \label{interpretations:surfaces:bulkboundary} CBS describes the general decay of surface effects \cite{reuter-085412-2011, reuter-084707-2013}, which helps explore quantum size effects. As depicted in Figure \ref{fig:pdos}, the diagonal blocks of the GF (which provide access to the projected DOSs) are one way to investigate this behavior. From Eq.\ \eqref{eq:gf-blocks:diagonal}, diagonal blocks of the GF are easily calculated from surface GFs for systems with the appropriate numbers of layers. More specifically, $\mathbf{G}_{j,j}(E)$ requires the surface GFs for the system with all layers to the left of layer $j$ [$\widetilde{\mathbf{G}}_{j-1}^\mathrm{L}(E)$] and for the system with all layers to the right of layer $j$ [$\widetilde{\mathbf{G}}_{N-j}^\mathrm{R}(E)$]. We thus see that surface effects persist as long as these two surface GFs do not resemble the surface GFs for semi-infinite systems; the ``bulk'' material would see semi-infinite systems on both sides. Because point \ref{interpretations:surfaces:converge} describes how CBS determines the convergence of these surface GFs to their semi-infinite limits, the subsurface properties of a material are also tied to CBS.
\end{enumerate}

\subsection{Interfaces}
\label{sec:interpretations:interfaces}
Many of the CBS results for surfaces readily generalize to interfaces \cite{mostoller-552-1979, brasher-4868-1980, schulman-2346-1983}, where two materials come into contact with each other.

\begin{enumerate}[resume*=interpretations]
\item \label{interpretations:interfaces:levelalignment} CBS describes the band ``lineup'' between the materials on both sides of the interface when neither material is metallic \cite{tersoff-4874-1984, margaritondo-2526-1985, demkov-195306-2005, monch-1113724-2011}. In almost every case, the two materials will have different workfunctions, meaning that electrons will spontaneously flow from one material to the other when the interface forms. The Fermi energies of the ``donor'' and ``acceptor'' materials will locally fall and rise near the interface, respectively, until there is no more charge transfer across the interface. This results in an interface dipole. Tersoff \cite{tersoff-4874-1984} posited that charge is transferred from ``conduction-like'' states of the donor to ``valence-like'' states of the acceptor, implying that equilibration occurs when ``non-bonding'' states from both materials are the next to be used. Using point \ref{interpretations:general:character}, the band lineup is thus determined by the branch point energies of the materials. This idea seems to work in many, but not all, cases \cite{monch-1113724-2011}.

\item \label{interpretations:interfaces:migs} Metal-induced gap states (MIGSs) at metal-semiconductor interfaces arise from semiconductor states with complex $k$ \cite{tersoff-465-1984}. In the band gap, these states penetrate several layers into the metal, but decay according to $\mathrm{Im}(k)$. Using similar reasoning to point \ref{interpretations:interfaces:levelalignment}, the metal's Fermi energy is pinned at or near the branch point energy of the semiconductor. This interpretation has also been used to describe the level alignment between an oligomeric molecule and metal surfaces for molecular electronics applications \cite{tomfohr-245105-2002, tomfohr-59-2002, tomfohr-1542-2004, wang-016401-2004}, although it may not always be appropriate \cite{lee-215204-2007}.
\end{enumerate}

\section{Singular Coupling Operators}
\label{sec:singular}

Sections \ref{sec:cbs-wf} and \ref{sec:cbs-gf} both assumed that the interlayer coupling operator, $\mathbf{H}_\mathrm{S}$, was nonsingular and, subsequently, used $\mathbf{H}_\mathrm{S}^{-1}$ in deriving CBS. Analytically, $\mathbf{H}_\mathrm{S}$ is probably not singular---there is infinitesimally weak coupling---but, in practice, it may be poorly conditioned such that numerically calculating or using $\mathbf{H}_\mathrm{S}^{-1}$ is error-prone \cite{bk:trefethen-1997, bk:golub-2012}. More physically, $\mathrm{rank}(\mathbf{H}_\mathrm{S})$ can be roughly interpreted as the number of ``bonds'' between layers \cite{dwivedi-134304-2016} such that, for example, $\mathbf{H}_\mathrm{S}$ will be \mbox{(near-)singular} when the repeat unit is too big. The choice of layers in Figure \ref{fig:asymmetric} exemplifies such a case; as seen in Figure \ref{fig:asymmetric}(a), the $-\alpha$ orbitals in layer $n$ (white circles) do not directly couple to the $+\alpha$ orbitals in layer $n+1$ (red circles).

Such \mbox{(near-)singularity} of $\mathbf{H}_\mathrm{S}$ has an immediate physical interpretation that will be helpful below. To see this, note that the transfer matrix $\mathbf{T}$ tends to singularity as $\mathbf{H}_\mathrm{S}$ tends to singularity (\textit{vide infra}). Then, because an eigenvalue $\lambda$ of the transfer matrix is related to a (complex) wavevector by $\lambda=e^{ika}$, (near-)singularity of $\mathbf{T}$ (\textit{i.e.},\ $\lambda\simeq0$) implies the existence of Bloch states with $\mathrm{Im}(k)\gg0$. Point \ref{properties:1d:pairedevals} further showed that $1/\lambda^\ast$ will be an eigenvalue of $\mathbf{T}$, meaning that there is also a state with $\mathrm{Im}(k)\ll 0$. These highly evanescent states rapidly decay (grow) from one layer to the next and are not likely to be physically important because of their strong localization \cite{osbourn-2124-1979, ram-mohan-6151-1988, boykin-8107-1996, mavropoulos-1088-2000, khomyakov-195402-2004, sorensen-155301-2008, sorensen-205322-2009, vergniory-544-2010, reuter-085412-2011}. It is possible that they can be approximated or even neglected.

In the end, there are several good reasons why the layers may be chosen such that $\mathbf{H}_\mathrm{S}$ is \mbox{(near-)singular}. For example, such a layer spacing could be the smallest that (i) requires only nearest-neighbor coupling between layers and (ii) makes all layers identical. These conditions correspond to the block tridiagonal and block Toeplitz structures of $\mathbf{H}$, respectively. Fortunately, several strategies have been developed for addressing the \mbox{(near-)singularity} of $\mathbf{H}_\mathrm{S}$ when formulating CBS \cite{chang-3975-1982, biczo-51-1985, mailhiot-8360-1986, brand-607-1987, bowen-2754-1995, boykin-7670-1996, boykin-8107-1996, hjort-5245-2000, tomfohr-245105-2002, fagas-268-2004, khomyakov-195402-2004, velev-r637-2004, li-194113-2006, rungger-035407-2008, reuter-085412-2011, dwivedi-134304-2016}. We discuss four in turn.

\subsection{Choose Smaller Layers}
\label{sec:singular:smaller-layers}
If large layers cause $\mathbf{H}_\mathrm{S}$ to be (near-)singular (see our example in Figure \ref{fig:asymmetric}), an obvious solution is to choose smaller layers. The problem is that smaller layers could lead to deviations from the block tridiagonal and block Toeplitz structure of the Hamiltonian, which were essential to deriving CBS in sections \ref{sec:cbs-wf} and \ref{sec:cbs-gf}. We defer our discussion of going beyond the block tridiagonal structure to section \ref{sec:extended-coupling} and focus here on relaxing the block Toeplitz structure.

In such a case, the Hamiltonian would instead have a periodic block Toeplitz structure \cite{chang-3975-1982, mostoller-6168-1982, schulman-2346-1983, rozsa-35-1992}, where the blocks along each diagonal cycle through some finite number of distinct blocks. Such a system is essentially a superlattice. For example, there are two different small layers that alternate in Figure \ref{fig:asymmetric}; the Hamiltonian is structured as
\[
\mathbf{H} = \left[
\begin{array}{cc|cc|c}
\mathbf{H}_{\mathrm{D,1}} & \mathbf{H}_{\mathrm{S},1}^\dagger & \mathbf{0} & \mathbf{0} & \cdots \\
\mathbf{H}_{\mathrm{S},1} & \mathbf{H}_{\mathrm{D},2} & \mathbf{H}_{\mathrm{S},2}^\dagger & \mathbf{0} & \cdots \\ \hline
\mathbf{0} & \mathbf{H}_{\mathrm{S},2} & \mathbf{H}_{\mathrm{D},1} & \mathbf{H}_{\mathrm{S},1}^\dagger & \cdots \\
\mathbf{0} & \mathbf{0} & \mathbf{H}_{\mathrm{S},1} & \mathbf{H}_{\mathrm{D},2} & \cdots \\ \hline
\vdots & \vdots & \vdots & \vdots & \ddots \\
\end{array}
\right],
\]
where the lines show that the block Toeplitz structure is restored when two small layers are grouped together. We can address this structure by defining two transfer matrices (similarly for MMTs): One moves from a layer of type 1 to a layer of type 2 and \textit{vice versa}. Using the associativity of matrix multiplication (MMT action), we can produce an ``aggregate'' transfer matrix (MMT) for the periodic structure \cite{schulman-2346-1983, ghahramani-1102-1989, reuter-084707-2013}.

\subsection{Approximate the Highly Evanescent States}
\label{sec:singular:approximate}
We can approximate the very evanescent states, which lead to the \mbox{(near-)singularity}, by states that are less evanescent \cite{mailhiot-8360-1986, brand-607-1987, rungger-035407-2008, reuter-085412-2011}. As long as the new states still decay (grow) much more rapidly than the physically-important states, the error can be negligible (or at least controllable). In essence, we compute the singular value decomposition \cite{bk:strang-2006} of $\mathbf{H}_\mathrm{S}$ and artificially increase the small singular values. This results in an approximate coupling operator $\overline{\mathbf{H}}_\mathrm{S}$ that has a condition number determined by the amount of increase. $\overline{\mathbf{H}}_\mathrm{S}$ is thus invertible. The derivations in sections \ref{sec:cbs-wf} and \ref{sec:cbs-gf} are then readily applicable using $\overline{\mathbf{H}}_\mathrm{S}$ instead of $\mathbf{H}_\mathrm{S}$.

\subsection{Use a Generalized Eigenvalue Problem}
\label{sec:singular:geneig}
The invertibility of $\mathbf{H}_\mathrm{S}$ was not needed until the final steps of deriving the transfer matrix [between Eqs.\ \eqref{eq:schrodinger-recurrence} and \eqref{eq:supercell-recurrence}]. Instead, we can rearrange Eq.\ \eqref{eq:schrodinger-recurrence} to obtain
\[
\mathbf{H}_\mathrm{S}^\dagger \ket{\psi_{n+1}} = (E\mathbf{I}-\mathbf{H}_\mathrm{D})\ket{\psi_n} - \mathbf{H}_\mathrm{S}\ket{\psi_{n-1}},
\]
where we stop just before inverting $\mathbf{H}_\mathrm{S}^\dagger$. As in section \ref{sec:cbs-wf}, we combine this equation with the tautology $\ket{\psi_n}=\ket{\psi_n}$ to produce \cite{biczo-51-1985, boykin-7670-1996, boykin-8107-1996, fagas-268-2004, khomyakov-195402-2004, velev-r637-2004, li-194113-2006, rungger-035407-2008}
\[
\left[ \begin{array}{c} \mathbf{H}_\mathrm{S}^\dagger \ket{\psi_{n+1}} \\ \ket{\psi_n} \end{array} \right] = \left[ \begin{array}{cc} E\mathbf{I}-\mathbf{H}_\mathrm{D} & -\mathbf{H}_\mathrm{S} \\ \mathbf{I} & \mathbf{0} \end{array} \right] \left[ \begin{array}{c} \ket{\psi_n} \\ \ket{\psi_{n-1}} \end{array} \right].
\]
Converting to the ``supercell'' wavefunction,
\[
\left[ \begin{array}{cc} \mathbf{H}_\mathrm{S}^\dagger & \mathbf{0} \\ \mathbf{0} & \mathbf{I} \end{array} \right] \ket{\Psi_{n+1}} = \left[ \begin{array}{cc} E\mathbf{I} - \mathbf{H}_\mathrm{D} & -\mathbf{H}_\mathrm{S} \\ \mathbf{I} & \mathbf{0} \end{array} \right] \ket{\Psi_{n}},
\]
and invoking Bloch's theorem [Eq.\ \eqref{eq:bloch-theorem}], we get
\begin{equation}
e^{ika} \left[ \begin{array}{cc} \mathbf{H}_\mathrm{S}^\dagger & \mathbf{0} \\ \mathbf{0} & \mathbf{I} \end{array} \right] \ket{\Psi_{n}} = \left[ \begin{array}{cc} E\mathbf{I} - \mathbf{H}_\mathrm{D} & -\mathbf{H}_\mathrm{S} \\ \mathbf{I} & \mathbf{0} \end{array} \right] \ket{\Psi_{n}}.
\label{eq:wf-gen-eigval}
\end{equation}
We thus avoid inverting $\mathbf{H}_\mathrm{S}^{\dagger}$ at the expense of having to solve a generalized eigenvalue problem \cite{bk:golub-2012} for CBS.

Generalized eigenvalue problems are more sophisticated than standard eigenvalue problems in that they admit both finite and infinite eigenvalues. For the present discussion, the generalized eigenvalue problem has a zero or infinite eigenvalue when
\[
\left[ \begin{array}{cc} E\mathbf{I} - \mathbf{H}_\mathrm{D} & -\mathbf{H}_\mathrm{S} \\ \mathbf{I} & \mathbf{0} \end{array} \right] \text{ or } \left[ \begin{array}{cc} \mathbf{H}_\mathrm{S}^\dagger & \mathbf{0} \\ \mathbf{0} & \mathbf{I} \end{array} \right]
\]
is singular, respectively \cite{bk:golub-2012}. Both of these cases arise when $\mathbf{H}_\mathrm{S}$ is singular, giving our justification that $\mathbf{T}$ tends to singularity as $\mathbf{H}_\mathrm{S}$ becomes singular. Note that, when $\mathbf{H}_\mathrm{S}$ is near-singular (such that the transfer matrix is analytically well-defined), $\mathbf{T}$ will have very large (in magnitude) eigenvalues that correspond to the nearly infinite eigenvalues of Eq.\ \eqref{eq:wf-gen-eigval}. In practice, these eigenvalues may lead to numerical instabilities when applying CBS \cite{mailhiot-8360-1986, ko-9945-1988, ting-985-1999}, which partly motivates the next strategy.

\subsection{Exclude the Highly Evanescent States}
\label{sec:singular:exclude}
Finally, we can build on the previous idea and remove the 0 and infinite eigenvalues. In essence, we deflate \cite{bk:trefethen-1997} the generalized eigenvalue problem so that only the states of interest remain \cite{brand-607-1987, khomyakov-195402-2004, rungger-035407-2008, sorensen-155301-2008, sorensen-205322-2009, dwivedi-134304-2016} by projecting the highly evanescent states out of the problem. Unlike in section \ref{sec:singular:approximate}, these states are not approximated; rather, they are completely removed from consideration. After deflation, the resulting (and approximate) $\mathbf{H}_\mathrm{S}$ is invertible such that the tools in sections \ref{sec:cbs-wf} and \ref{sec:cbs-gf} are again applicable. Compared to the generalized eigenvalue problem in section \ref{sec:singular:geneig}, this procedure requires a bit of extra numerical effort to perform the deflation, but ultimately produces smaller matrices for the CBS computations.

\section{Extended Coupling Between Layers}
\label{sec:extended-coupling}

As discussed in the previous section, \mbox{(near-)singularity} of the inter-layer coupling operator $\mathbf{H}_\mathrm{S}$ may prompt us to choose smaller layers, which may, in turn, break the block tridiagonal structure of $\mathbf{H}$. This is but one reason why the block tridiagonal structure may be too restrictive, and in this section we discuss extended couplings. In a broad sense, a block banded matrix (which includes block tridiagonal and block pentadiagonal matrices, \textit{etc}.)\ remains block quasi-separable, meaning all of the ideas about generators---and CBS---are still applicable. As we now see, extended couplings do not fundamentally change anything; they only require more complicated algebra.

As an illustrative example, we first consider a system with nearest-neighbor and next-nearest-neighbor couplings before discussing the general case. Accordingly, the Hamiltonian becomes block pentadiagonal (and is still block Toeplitz),
\[
\mathbf{H} = \left[
\begin{array}{cccccc}
\mathbf{H}_\mathrm{D} & \mathbf{H}_{\mathrm{S},1}^\dagger & \mathbf{H}_{\mathrm{S},2}^\dagger & \mathbf{0} & \mathbf{0} & \cdots \\
\mathbf{H}_{\mathrm{S},1} & \mathbf{H}_\mathrm{D} & \mathbf{H}_{\mathrm{S},1}^\dagger & \mathbf{H}_{\mathrm{S},2}^\dagger & \mathbf{0} & \cdots \\
\mathbf{H}_{\mathrm{S},2} & \mathbf{H}_{\mathrm{S},1} & \mathbf{H}_\mathrm{D} & \mathbf{H}_{\mathrm{S},1}^\dagger & \mathbf{H}_{\mathrm{S},2}^\dagger & \cdots \\
\mathbf{0} & \mathbf{H}_{\mathrm{S},2} & \mathbf{H}_{\mathrm{S},1} & \mathbf{H}_\mathrm{D} & \mathbf{H}_{\mathrm{S},1}^\dagger & \cdots \\
\mathbf{0} & \mathbf{0} & \mathbf{H}_{\mathrm{S},2} & \mathbf{H}_{\mathrm{S},1} & \mathbf{H}_\mathrm{D} & \cdots \\
\vdots & \vdots & \vdots & \vdots & \vdots & \ddots \\
\end{array}
\right].
\]
We will assume $\mathbf{H}_{\mathrm{S},2}$ is nonsingular, noting that all of the ideas discussed in section \ref{sec:singular} are generalizable if it is not.

Let us first focus on diagonalizing $\mathbf{H}$. Following section \ref{sec:cbs-wf}, we can derive a similar statement to Eq.\ \eqref{eq:schrodinger-recurrence},
\[
E\ket{\psi_n} = \mathbf{H}_{\mathrm{S},2} \ket{\psi_{n-2}} + \mathbf{H}_{\mathrm{S},1} \ket{\psi_{n-1}} + \mathbf{H}_\mathrm{D} \ket{\psi_n} + \mathbf{H}_{\mathrm{S},1}^\dagger \ket{\psi_{n+1}} + \mathbf{H}_{\mathrm{S},2}^\dagger \ket{\psi_{n+2}}.
\]
This equation, along with a tautology for each layer ($\ket{\psi_j}=\ket{\psi_j}$), leads to
\begin{equation}
\left[ \begin{array}{c}
\ket{\psi_{n+2}} \\
\ket{\psi_{n+1}} \\
\ket{\psi_n} \\
\ket{\psi_{n-1}} \\
\end{array} \right] = \left[ \begin{array}{cccc}
-\mathbf{H}_{\mathrm{S},2}^{-\dagger} \mathbf{H}_{\mathrm{S},1}^\dagger & -\mathbf{H}_{\mathrm{S},2}^{-\dagger} (E\mathbf{I} - \mathbf{H}_\mathrm{D}) & -\mathbf{H}_{\mathrm{S},2}^{-\dagger} \mathbf{H}_{\mathrm{S},1} & -\mathbf{H}_{\mathrm{S},2}^{-\dagger} \mathbf{H}_{\mathrm{S},2} \\
\mathbf{I} & \mathbf{0} & \mathbf{0} & \mathbf{0} \\
\mathbf{0} & \mathbf{I} & \mathbf{0} & \mathbf{0} \\
\mathbf{0} & \mathbf{0} & \mathbf{I} & \mathbf{0} \\
\end{array} \right] \left[ \begin{array}{c}
\ket{\psi_{n+1}} \\
\ket{\psi_n} \\
\ket{\psi_{n-1}} \\
\ket{\psi_{n-2}} \\
\end{array} \right].
\label{eq:pentadiagonal-recurrence}
\end{equation}
The matrix on the right-hand side is essentially the companion matrix of Ref.\ \onlinecite{chang-3975-1982}, which is a ``multi-layer'' generalization of the transfer matrix. We denote this companion matrix by $\mathbf{C}$ in what follows. We can also generalize the supercell wavefunction by making it larger,
\[
\ket{\Psi_{n+1}} \equiv \left[ \begin{array}{c}
\ket{\psi_{n+2}} \\
\ket{\psi_{n+1}} \\
\ket{\psi_n} \\
\ket{\psi_{n-1}} \\
\end{array} \right],
\]
such that Eq.\ \eqref{eq:pentadiagonal-recurrence} becomes
\[
\ket{\Psi_{n+1}} = \mathbf{C} \ket{\Psi_{n}}.
\]
Finally, Bloch's theorem again specifies that $\ket{\Psi_{n+1}}=e^{ika}\ket{\Psi_{n}}$, such that
\begin{equation}
e^{ika} \ket{\Psi_{n}} = \mathbf{C} \ket{\Psi_{n}}.
\label{eq:companion}
\end{equation}
Mirroring the transfer matrix for a Hamiltonian with nearest-neighbor couplings, the eigenvalues of $\mathbf{C}$ produce the CBS for a Hamiltonian with extended couplings.

Turning now to the inversion problem (\textit{i.e.},\ calculating the GF), there are fewer direct results. We know of only one study that has explicitly investigated the inverses of block pentadiagonal matrices \cite{koulaei-223-2007}, and it derived recurrence relations for the blocks of the inverse that are similar to Eq.\ \eqref{eq:gf-blocks}. That said, and as one might expect, these recurrence relations are more complicated. It is not immediately obvious how to access CBS from them when the block Toeplitz structure is also present. More work needs to be performed to derive generators for the inverses of block pentadiagonal and block Toeplitz matrices.

Generalizing to higher-order extended couplings, the Hamiltonian becomes block banded and block Toeplitz. For example, if a layer couples to its nearest $C$ neighbors on either side,
\[
\mathbf{H} = \left[
\begin{array}{cccccccc}
\mathbf{H}_\mathrm{D} & \mathbf{H}_{\mathrm{S},1}^\dagger & \mathbf{H}_{\mathrm{S},2}^\dagger & \cdots & \mathbf{H}_{\mathrm{S},C}^\dagger & \mathbf{0} & \mathbf{0} & \cdots \\
\mathbf{H}_{\mathrm{S},1} & \mathbf{H}_\mathrm{D} & \mathbf{H}_{\mathrm{S},1}^\dagger & \cdots & \mathbf{H}_{\mathrm{S},C-1}^\dagger & \mathbf{H}_{\mathrm{S},C}^\dagger & \mathbf{0} & \cdots \\
\mathbf{H}_{\mathrm{S},2} & \mathbf{H}_{\mathrm{S},1} & \mathbf{H}_\mathrm{D} & \cdots & \mathbf{H}_{\mathrm{S},C-2}^\dagger & \mathbf{H}_{\mathrm{S},C-1}^\dagger & \mathbf{H}_{\mathrm{S},C}^\dagger & \cdots \\
\vdots & \vdots & \vdots & \ddots & \vdots & \vdots & \vdots \\
\mathbf{H}_{\mathrm{S},C} & \mathbf{H}_{\mathrm{S},C-1} & \mathbf{H}_{\mathrm{S},C-2} & \cdots & \mathbf{H}_\mathrm{D} & \mathbf{H}_{\mathrm{S},1}^\dagger & \mathbf{H}_{\mathrm{S},2}^\dagger & \cdots \\
\mathbf{0} & \mathbf{H}_{\mathrm{S},C} & \mathbf{H}_{\mathrm{S},C-1} & \cdots & \mathbf{H}_{\mathrm{S},1} & \mathbf{H}_\mathrm{D} & \mathbf{H}_{\mathrm{S},1}^\dagger & \cdots \\
\mathbf{0} & \mathbf{0} & \mathbf{H}_{\mathrm{S},C} & \cdots & \mathbf{H}_{\mathrm{S},2} & \mathbf{H}_{\mathrm{S},2} & \mathbf{H}_\mathrm{D} & \cdots \\
\vdots & \vdots & \vdots & & \vdots & \vdots & \vdots & \ddots \\
\end{array}
\right].
\]
As above, we assume $\mathbf{H}_{\mathrm{S},C}$ is nonsingular, again referring the reader to the techniques developed in section \ref{sec:singular} if it is not. From this structure of $\mathbf{H}$, we can similarly derive the companion matrix
\[
\mathbf{C} = \left[ \begin{array}{ccccccc}
-\mathbf{H}_{\mathrm{S},C}^{-\dagger} \mathbf{H}_{\mathrm{S},C-1}^\dagger & -\mathbf{H}_{\mathrm{S},C}^{-\dagger} \mathbf{H}_{\mathrm{S},C-2}^\dagger & \cdots &\mathbf{H}_{\mathrm{S},C}^{-\dagger} (E\mathbf{I} - \mathbf{H}_\mathrm{D}) & \cdots & -\mathbf{H}_{\mathrm{S},C}^{-\dagger} \mathbf{H}_{\mathrm{S},C-1} & -\mathbf{H}_{\mathrm{S},C}^{-\dagger} \mathbf{H}_{\mathrm{S},C} \\
\mathbf{I} & \mathbf{0} & & & & \\
& \mathbf{I} & \ddots & & & \\
& & \ddots & \mathbf{0} & & \\
& & & \mathbf{I} & \ddots & \\
& & & & \ddots & \mathbf{0} \\
& & & & & \mathbf{I} & \mathbf{0} \\
\end{array} \right],
\]
where omitted blocks are $\mathbf{0}$ (after the first block row, only the block diagonal and block sub-diagonal are shown). Using an expanded supercell wavefunction,
\[
\ket{\Psi_{n+1}} \equiv \left[ \begin{array}{c}
\ket{\psi_{n+C}} \\
\ket{\psi_{n+C-1}} \\
\vdots \\
\ket{\psi_{n-C+1}} \\
\end{array} \right],
\]
we can again write $\ket{\Psi_{n+1}} = \mathbf{C} \ket{\Psi_{n}}$, which becomes Eq.\ \eqref{eq:companion} with the application of Bloch's theorem. Analogously, the eigenvalues of our generalized $\mathbf{C}$ still access the material's CBS. Regarding the GF, algorithms for inverting block banded matrices have been developed and discussed in the mathematical literature \cite{rozsa-447-1989, bk:vandebril-2008-v1}, but do not provide such a clear connection to CBS when the block Toeplitz structure is also imposed. We know of one work \cite{bini-431-1988} that specifically considers the inverses of block banded and block Toeplitz matrices. As in the block pentadiagonal case, future investigations into this problem are needed to determine a precise relation between CBS and the GF.

\section{Conclusions}
\label{sec:conclusions}
Complex band structure is an intrinsic material property for describing the physics of materials that have a prevalent repeat unit but lack perfect translational symmetry. Although it was first developed to answer fundamental questions regarding the implications of complex-valued wavevectors in Bloch's theorem, it subsequently found numerous applications throughout condensed matter physics, ranging from surfaces and interfaces to topological materials. The development of CBS, however, has generally been limited by, and sometimes in spite of, this broad applicability. Many studies independently redeveloped the main concepts and results of CBS, often with little comparison or connection to previous efforts. As a result, there are several seemingly disparate formulations of CBS in the literature.

Our primary objective in this work was to develop the interpretation that CBS is the minimal information that describes all of a material's static and dynamic electronic properties. This result is a mathematical implication of the block tridiagonal and block Toeplitz structure of a one-dimensional material's Hamiltonian (with generalizations discussed in sections \ref{sec:singular} and \ref{sec:extended-coupling}). In this way, each formulation of CBS taps into this intrinsic material information to address a specific physical problem.

Using this understanding of CBS and a consistent notation, we proceeded to unify many of CBS's reported properties and applications (sections \ref{sec:properties} and \ref{sec:interpretations}). Instead of viewing them as implications of CBS, we showed that they are intrinsic material properties and inherently described by CBS. In this way, our discussion showcased the applicability and consistency of the various CBS formulations. It also occurred at the level of operators, not matrices, to obviate unnecessary discussions of basis-set effects. We ultimately hope that this work provides an accessible introduction to CBS, outlines its utlity, and encourages its further development.

Analyzing CBS in terms of the Hamiltonian's information content exposes two potential future directions. (i) Higher-dimensional CBS, where $\vec{k}$ has complex components in more than one dimension, needs to be studied in more detail. In the absence of impurities (for example), each unit cell will have the same diagonal block ($\mathbf{H}_\mathrm{D}$) and will couple to its neighbors with a direction-dependent element (generalizing $\mathbf{H}_\mathrm{S}$). Collectively, this is still $\mathcal{O}(1)$ information; the key difference is that the material's topology is no longer linear. CBS is conceptually unchanged, and we need only build on fundamental discussions by Blount \cite{blount-305-1962} and Heine \cite{heine-300-1963} to better access CBS. Along these lines, some non-linear material topologies have recently been investigated, including Cayley trees \cite{jiang-057202-2012} and cylinders \cite{nguyen-275301-2016}. (ii) Adding impurities to the Hamiltonian, where the information becomes $\mathcal{O}(n)$ in the number of defects, would also be an interesting endeavor. CBS alone will no longer fully describe the defected material's electronic structure, but there should be insightful and relatively inexpensive techniques to incorporate the disorder. We recently reported \cite{reuter-014009-2012} some mathematical preliminaries on the topic.

\begin{acknowledgments}
I am grateful to Caroline Taylor, Thorsten Hansen, Jay Bardhan, Mark Ratner, and Jefferson Bates for helpful conversations. This research was supported by startup funds from the Institute for Advanced Computational Science at Stony Brook University. The figures were prepared using the SciDraw package \cite{caprio-107-2005}.
\end{acknowledgments}

\bibliographystyle{apsrev4-1}
\bibliography{/Users/mgreuter/Documents/Publications/library,notes}

\end{document}